\documentclass[reprint, twocolumn, amsmath, amssymb, superscriptaddress]{revtex4-2}

\usepackage{graphicx}
\usepackage{dcolumn}
\usepackage{amsmath}

\usepackage[usenames,dvipsnames]{xcolor} 
\usepackage{hhline}
\usepackage{array}
\newcolumntype{P}[1]{>{\centering\arraybackslash}p{#1}}
\usepackage{textcomp}
\usepackage[normalem]{ulem}
\usepackage{changes}

\begin{document}

\title{In-situ tuning of cavity dissipation and a topological transition in an atom-nanotip-cavity system} 
\author{Taegyu Ha}
\thanks{These authors contributed equally to this work.}
\affiliation{Department of Electrical Engineering, Pohang University of Science and Technology (POSTECH), 37673 Pohang, Korea}
\author{Kiyanoush Goudarzi$^{\dagger ,}$} 
\thanks{These authors contributed equally to this work.}
\affiliation{Department of Electrical Engineering, Pohang University of Science and Technology (POSTECH), 37673 Pohang, Korea}
\author{Dowon Lee}
\affiliation{Department of Electrical Engineering, Pohang University of Science and Technology (POSTECH), 37673 Pohang, Korea}
\author{Donggeon Kim}
\affiliation{Department of Electrical Engineering, Pohang University of Science and Technology (POSTECH), 37673 Pohang, Korea}
\author{Eunchul Jeong}
\affiliation{Graduate School of Convergence Science and Technology, Pohang University of Science and Technology (POSTECH), 37673 Pohang, Korea}
\author{Uijin Kim}
\affiliation{Department of Electrical Engineering, Pohang University of Science and Technology (POSTECH), 37673 Pohang, Korea}
\author{Myunghun Kim}
\affiliation{Department of Electrical Engineering, Pohang University of Science and Technology (POSTECH), 37673 Pohang, Korea}
\author{Jinuk Kim}
\affiliation{Quantum Technology Institute, Korea Research Institute of Standards and Science (KRISS), 34113 Daejeon, Korea}
\author{Moonjoo Lee}
\email{kgoudarzi@postech.ac.kr $and$ moonjoo.lee@postech.ac.kr}
\affiliation{Department of Electrical Engineering, Pohang University of Science and Technology (POSTECH), 37673 Pohang, Korea}

\date{\today}

\begin{abstract}
We theoretically investigate a method for tuning a dissipation rate in an atom-cavity system.
By inserting a nanotip into the cavity mode, we estimate that the cavity dissipation rate increases by a factor of approximately $20$, due to the scattering loss at the tip.
As applications of our in-situ technique, we demonstrate that an exceptional line can be obtained when a single atom or ion is coupled to the resonator. 
Moreover, the position of an exceptional point is tuned by adjusting the decay rate, enabling a topological transition through dissipation control alone.  
\end{abstract}

\maketitle


Dissipation lies at the heart of open quantum systems. 
The irreversible coupling between a system and a reservoir has been a central research theme over the past few decades, particularly from the perspectives of quantum noise and fluctuations~\cite{Gardiner04}, as well as quantum trajectory~\cite{Plenio1998, Carmichael2008}. 
While dissipation is often regarded as detrimental to coherent processes, the dissipative dynamics can be harnessed in useful manners in quantum science and technology.
For example, in cavity quantum electrodynamics (QED), cavity decay consists of the sole channel for utilizing photons in quantum communication~\cite{Kimble08a, Reiserer2015}. 
Also, the dissipation mediates the generation of atomic entanglement in a resonator~\cite{Plenio99, Kastoryano11, Su2014}, the motion of atoms can be manipulated through the cavity decay~\cite{Wolke2012}, and the dissipation can even assist the formation ultracold molecules in an optical cavity~\cite{Wellnitz2020}.
Recently, the combination of dissipation with coherent interactions has paved the way for studying non-Hermitian physics in cavity QED~\cite{Choi2010, Lu2018, Huang2022, Li2023, Kim2023a, Lee2024, Agarwal2024}.

If the dissipation rate can be tuned, it provides an additional control parameter for manipulating quantum states. 
In quantum simulator experiments, the effective decay rate of atomic transitions was adjusted by varying laser parameters, which enabled the exploration of novel quantum phases of matter~\cite{Ferri2021}, electron transfer processes in molecules~\cite{So2024}, and higher-order exceptional points (EPs)~\cite{Li2024}.
In superconducting circuits, tunable dissipation was realized by incorporating controllable dissipative elements into the circuits~\cite{Jones2013, Silveri2019}. 
This approach has led the observation of phenomena such as the broadband Lamb shift~\cite{Silveri2019}, the crossing over EPs~\cite{Partanen2019}, and rapid resonator initialization~\cite{Maurya2024}---all demonstrated in the microwave domain.
In contrast, in the optical regime of cavity QED, the dissipation rates are just fixed by the mirror reflectivities and have not yet been treated as a tunable parameter.

Here, we theoretically investigate a method for tuning the decay rate of an optical cavity. 
A metal nanotip is inserted transversely into a Fabry-Pérot resonator to perturb the intracavity field.
As the tip intrudes into the mode, the enhanced scattering loss increases the overall cavity decay rate. 
This intrusion also causes the redshift of the resonance frequency, which can be compensated by adjusting the cavity length. 
One application of our technique includes the emergence of an exceptional line in the parameter space of the atom-cavity coupling constant $g$ and cavity decay rate $\kappa$. 
We further demonstrate that the position of an EP on the Riemann surface shifts as $\kappa$ changes.
Finally, we show that varying the cavity dissipation leads to a topological transition in the eigenvalue braiding behavior, as the system encircles an EP in parameter space.



\begin{figure}[!t]	
		\includegraphics[width=1 \linewidth]{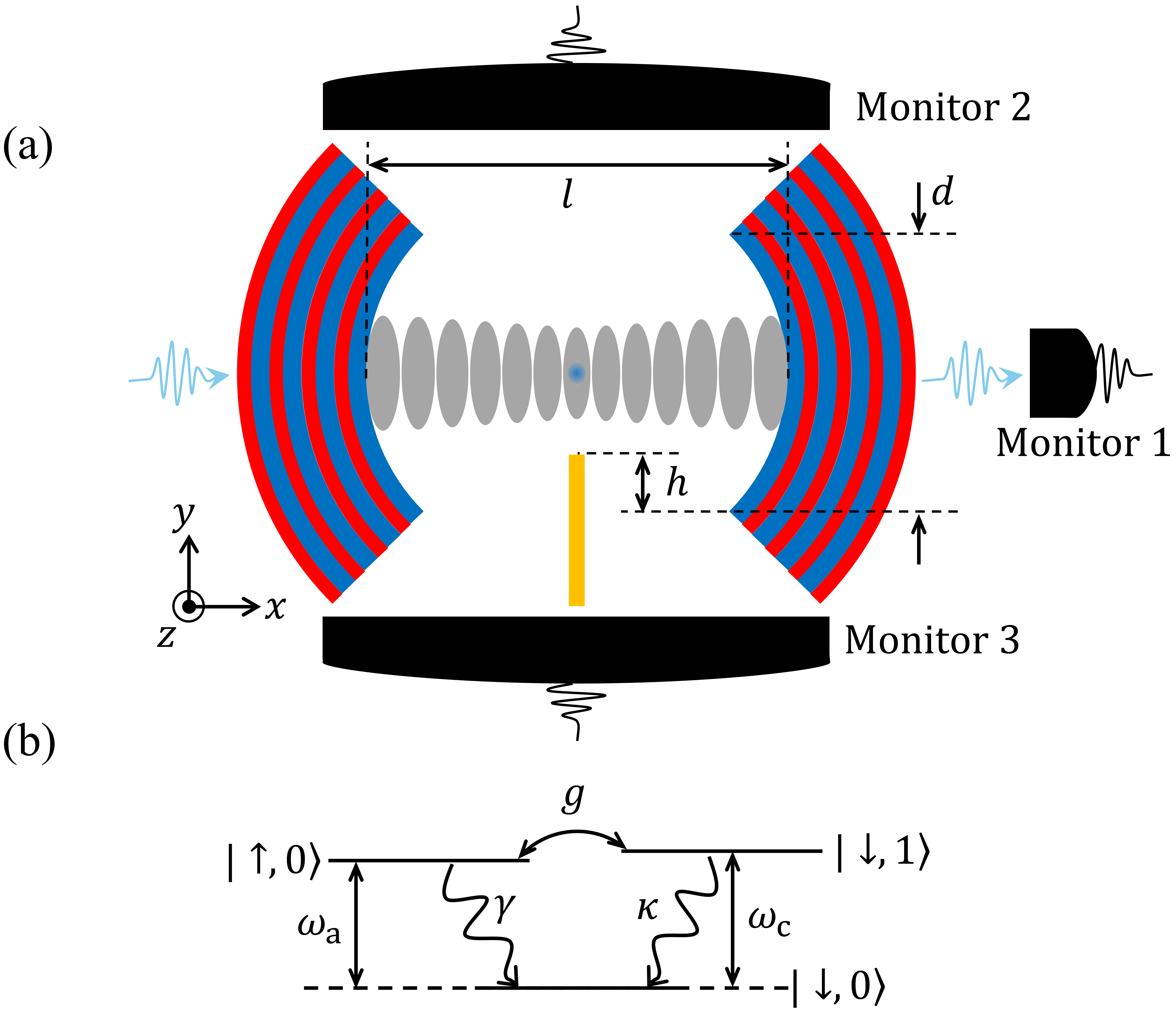}
		\caption{(a) Schematic of our atom-nanotip-cavity system.
			Cavity is driven with a weak probe field (light blue), and its transmission is measured using a photodetector (black).			
			The distance between the end of the nanotip (gold) and that of the substrate at the mirror facet is denoted by $h$.
			Coating layers of SiO$_{2}$ and Ta$_{2}$O$_{5}$ are shown in blue and red, respectively. 
			(b) Energy level diagram in the basis of atomic states and cavity photon numbers. 
			}	
	\label{fig:setup}
\end{figure}

Our atom-nanotip-cavity system consists of two mirrors that form a Fabry-P\'erot cavity, with a gold nanotip inserted from the bottom (Fig.~\ref{fig:setup}).  
The mirror coatings consist of 40 alternating layers of SiO$_{2}$ (refractive index of $1.4550$ and thickness of $0.134$~$\mu$m) and Ta$_{2}$O$_{5}$ ($2.0411$ and $0.095$~$\mu$m), and each mirror has a radius of curvature of $30~\mu$m.
The mirrors are separated by a distance of $l = 10.150~\mu$m and have a inner diameter of $d=20.500~\mu$m.
A gold nanotip with a width of $w = 300$~nm is positioned in the cavity.
We consider a single $^{87}\mathrm{Rb}$ atom is trapped in the cavity, where $g$ is controlled by exploiting the spatial variation of the cavity field~\cite{Lee2014}. 
One can change $g$ by, e.~g., moving optical tweezers, illuminated from a transverse direction, along the cavity axis.
The cavity resonance frequency is tuned close to the D$_2$ transition of $5^2S_{1/2} |F=2\rangle \leftrightarrow 5^2 P_{3/2} |F^{\prime}=3\rangle$ at 780 nm, which we denoted as \( |g\rangle \leftrightarrow |e\rangle \).
While our eigenvalue calculations are based on the parameters described above, the proposed setup is also adaptable to trapped ions. 
In that case, different mirror coatings are used, and the ion position can be varied via dc voltages applied to the trap electrodes~\cite{Guthoehrlein01}.

\begin{figure}[!t]
		\includegraphics[width=1 \linewidth]{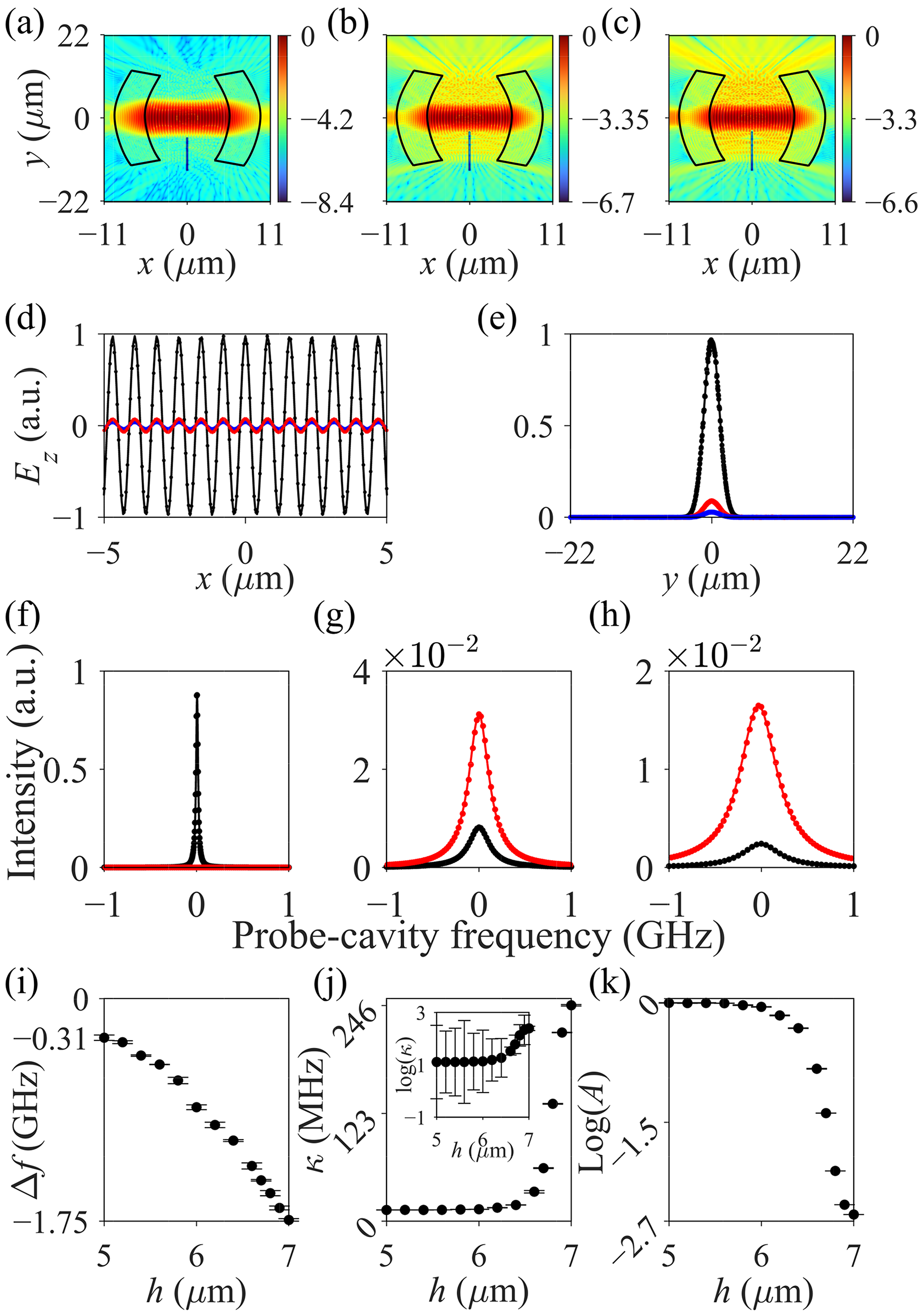}
		\caption{		
			Electric field distribution of the cavity mode at (a) $h=5$, (b) $6.8$, and (c) $7$~$\mu$m.			
			Color bar indicates $\log(|E_z|)$, where $E_z$ represents the normalized field.						
			Cross-sectional profiles of the electric field along the (d) $x$ and (e) $y$ axis, respectively, for $h=5$ (black), $6.8$ (red), and $7$~$\mu$m (blue). 
			Points correspond to the calculation results, and the lines represent fitted curves.
			(f)--(h) Calculated spectra as a function of probe-cavity frequency (black points at Monitor $1$, red points at Monitor $2$) for the corresponding values of $h$ in (a)--(c).
			Line indicates the fitting result to a Lorentzian function.
			 Detected signal at Monitor $3$ is on the order of $10^{-4}$ maximally.
			(i) Cavity resonance frequency shift, (j) cavity decay rate $\kappa$, and (k) the logarithm of transmission amplitude $A$, at Monitor $1$, as a function of $h$. 	
			Error bars represent uncertainties of the numerical calculation. 
			}
	\label{fig:field_calculation} 
\end{figure}


We describe calculations of the cavity resonance frequency shift and the change in $\kappa$ as the nanotip is introduced into the cavity. 
Fig.~\ref{fig:field_calculation} presents the electric field distribution in the cavity and analysis results  for $h=5\sim7$~$\mu$m, respectively. 
Two key features are observed in this configuration.
First, inserting the nanotip into the cavity gives rise to an increase of the effective refractive index of the nanotip-cavity system, leading to a redshift in the cavity's resonance frequency. 
Second, the perturbation of the cavity field leads to enhanced scattering loss.  
As illustrated in Figs.~\ref{fig:field_calculation}(a)--(c), the scattered field at the nanotip escapes the cavity mode prominently along the upward direction of the resonator.
Although the cavity field appears distorted in regions away from the center, in the central region it remains well-defined cosine function along the $x$ axis and a Gaussian profile, with a waist of $w_{0} = 1.70(6)$~$\mu$m, along the $y$ axis (Figs.~\ref{fig:field_calculation}(d)--(e)).
Thus the atom effectively interacts only with the fundamental mode, while its coupling to higher-order field components is negligible.

\begin{figure*}[!t]
		\includegraphics[width=1 \linewidth]{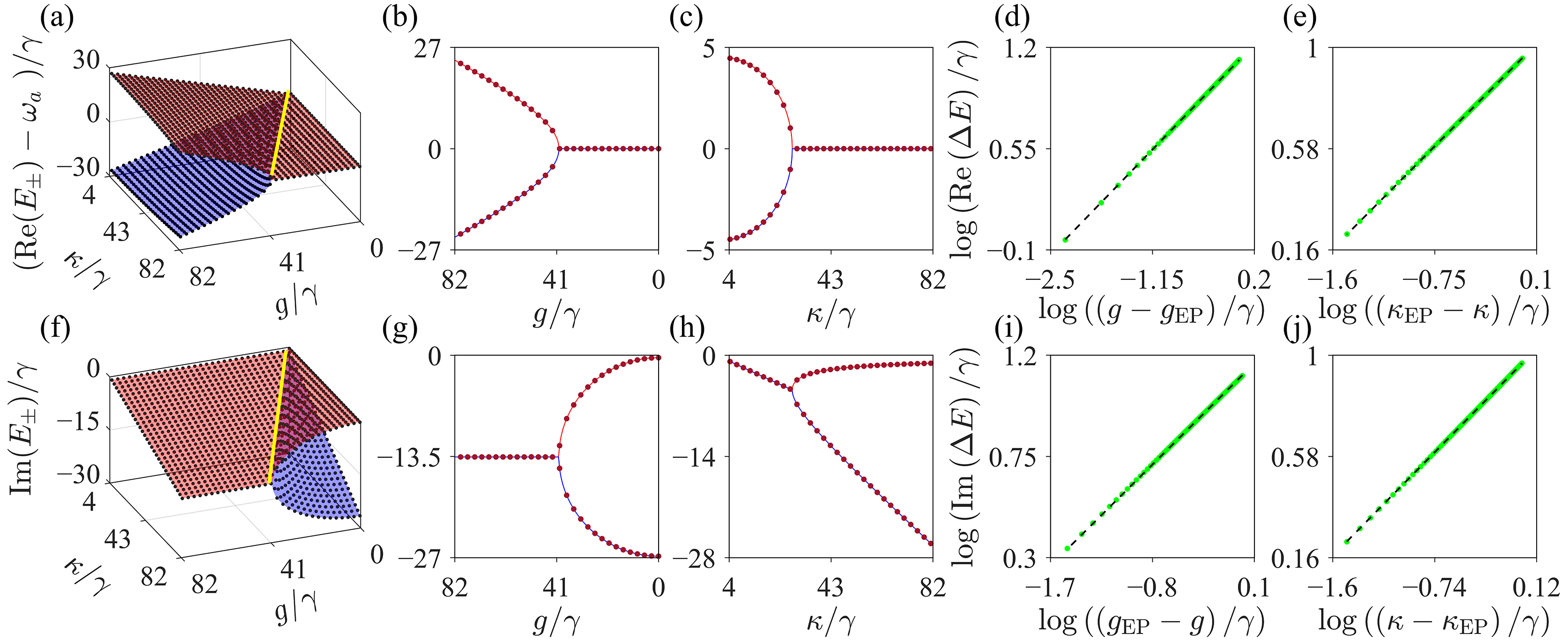}
		\caption{
			(a) Real part of the eigenvalues, $\mathrm{Re}(E_{\pm}) - \omega_{\textrm{a}}$, as functions of $g$ and $\kappa$.
			All values are normalized with $\gamma$, and $\omega_{\textrm{a}} = \omega_{\textrm{c}}$.
			Red and blue surfaces correspond to $E_{+}$ and $E_{-}$, respectively.  
			Exceptional line is represented in yellow.
			Results of QMC simulation are shown with brown points.
			Cross sectional view of $\mathrm{Re}(E_{\pm}) - \omega_{\textrm{a}}$ as a function of (b) $g/\gamma$ at $\kappa/\gamma = 81.2$, and (c) $\kappa/\gamma$ at $g/\gamma = 13.5$.			
			Log-log plot of $\textrm{Re} ( \Delta E ) / \gamma = \mathrm{Re}(E_{+} - E_{-})  / \gamma$ versus (d) $\left(g - g_{\mathrm{EP}}\right) / \gamma$ at $\kappa/\gamma = 81.2$, and (e) $\left( \kappa - \kappa_{\mathrm{EP}} \right) / \gamma$ at $g/\gamma = 13.5$, respectively.
			Green points are the differences of the eigenvalues, and black dashed line indicates the fitting result.  
			(f)--(j) Same plot as above for the imaginary part of the eigenvalues. 
		}
		\label{fig:exceptional_line}
\end{figure*}

In order to understand the phenomena more quantitatively, we fit the cavity transmission spectra at Monitor $1$ and intensities at Monitor $2$ and $3$ to Lorentzian functions (Figs.~\ref{fig:field_calculation}(f)--(h)). 
Note that, in Fig.~\ref{fig:field_calculation}(f), a reduced transmission of 0.88 at $\Delta_{\textrm{pc}} (=\omega_{\textrm{p}} - \omega_{\textrm{c}}) = 0$, where $\omega_\textrm{p}$ is the frequency of the probe laser and $\omega_{\textrm{c}}$ is the cavity resonance frequency, is determined by the scattering and absorption losses in the coating layers.
Each spectrum in the range of $h=0\sim7$~$\mu$m fits the model very well.
While the frequency changes mildly, from $\Delta f=0$ to $-0.31000(2)$~GHz for the tip positions from $h=0$ to $5$~$\mu$m, a notable shift of $1.44410(2)$~GHz occurs as the tip moves from $h = 5$ to $7$~$\mu$m.
In practice this frequency shift can be compensated by decreasing the cavity length, e.~g., using a piezoelectric transducer to stabilize the cavity resonance frequency, in such a way that the cavity frequency is stabilized to the atomic resonance frequency.
Regarding dissipation, while a slight increase of $\kappa/(2\pi) = 12.70(7)$ to $13.50(1)$~MHz is seen as the tip intrudes from $h=5$ to $6$~$\mu$m, the dissipation rate significantly increases to $\kappa/(2\pi)=245.00(5)$~MHz at $h=7$~$\mu$m.  
This means that we have a precise and broad control of $\kappa$: Nanoscopic positioning of the tip within the cavity mode, using a piezoelectric transducer that holds the tip, offers the tunability of dissipation rate. 


We then consider that a single atom is coupled to the optical cavity, in the presence of the nanotip. 
In the weak-excitation limit where the mean photon number $\langle n \rangle \ll 1$, we obtain the non-Hermitian Hamiltonian
\begin{equation}
	H_{\textrm{nH}} =
	\begin{pmatrix}
		\omega_{\textrm{a}} - i\gamma & g \\
		g & \omega_{\textrm{c}} - i\kappa
	\end{pmatrix}
	,
	\label{eq:H_nH}
\end{equation}

\noindent
where the atomic resonance frequency is $\omega_{\textrm{a}}$, atomic decay rate is $\gamma$, and $\hbar = 1$.
The eigenvalues ($E_\pm$) and eigenvectors ($|\pm\rangle$) of the Hamiltonian are given by
\begin{align}
		E_{\pm}	 & = g \left( B_{+} \pm \sqrt{B_{-}^{2} + 1} \right),  \\
		|+\rangle & = \cos \left( \theta \right) |\uparrow, 0\rangle + \sin \left( \theta \right) |\downarrow, 1\rangle, \\
		|-\rangle & = -\sin \left( \theta \right) |\uparrow, 0\rangle + \cos \left( \theta \right) |\downarrow, 1\rangle, 
\end{align}

\noindent 
with $B_{\pm} = \left( \omega_{\pm} - i \gamma_{\pm} \right) / g$, $\omega_{\pm} = (\omega_{a} \pm \omega_{c}) / 2$, $ \gamma_{\pm} = | \kappa \pm \gamma | /2$, and $\tan \left( \theta \right) = B_{-} + \sqrt{B_{-}^2 + 1}$. 
The atomic excited (ground) state is denoted by $\uparrow$ ($\downarrow$), and the eigenvectors satisfy the biorthogonality condition of $\langle +^{*} | + \rangle = \langle -^{*} | - \rangle$ and $\langle -^{*} | + \rangle = \langle +^{*} | - \rangle$.

\begin{figure*}[!t]
	\includegraphics[width=1\linewidth]{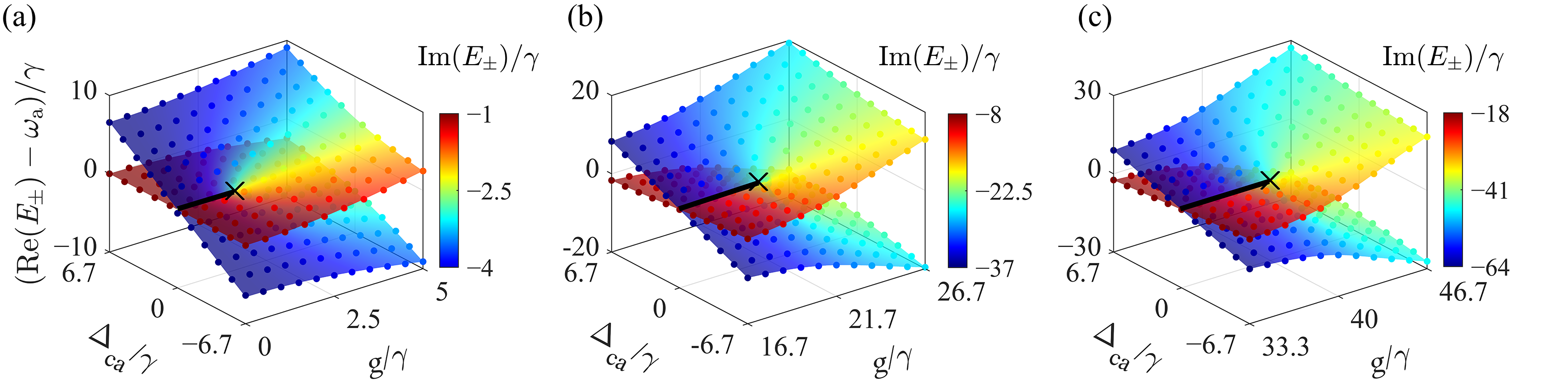}
	\caption{
		Riemann surface of the eigenvalues in the parameter space of $\Delta_{\textrm{ca}}/\gamma$ and $g/\gamma$, for (a) $h=5$~$\mu$m ($\kappa/(2\pi) = 12.7$~MHz), (b) $h=6.8$~$\mu$m ($\kappa/(2\pi) = 133$~MHz), and (c) $h=7$~$\mu$m ($\kappa/(2\pi) = 246$ MHz).  
		EP appears at $\Delta_{\textrm{ca}}/\gamma=0$, and $g/\gamma = 1.64$, $21.4$, and $40.1$, respectively. 
		Black line indicates the branch cut, and the EP is denoted by the cross.
		Points represent the results of QMC simulations.
		}
	\label{fig:Riemann_surface} 
\end{figure*}


Given fixed values of $g$ and $\kappa$, the eigenvalues coalesce to $E_{\pm} = \omega_{+} - i \gamma_{+}$ at the EP conditions of $\omega_{\textrm{a}} = \omega_{\textrm{c}}$ and $g = \gamma_{-}$. 
These two conditions can be simultaneously satisfied by placing the atom at a specific position within the cavity such that $g = \gamma_{-}$, while the cavity resonance frequency is tuned to the atomic resonance.
Now, suppose we vary both the atomic position and the nanotip location in such a way that the EP conditions are continuously maintained---this results in the emergence of an exceptional line in the parameter space of $g$ and $\kappa$.
Figs.~\ref{fig:exceptional_line}(a) and (f) display two surfaces corresponding to $E_{+}$ and $E_{-}$, respectively.
Those two surfaces coalesce at a line shown in yellow, which corresponds to the exceptional line. 
Cross-sectional views are shown for a given $g$ (Figs.~\ref{fig:exceptional_line}(b) and (g)) and $\kappa$ (Figs.~\ref{fig:exceptional_line}(c) and (h)). 
As presented in Figs.~\ref{fig:exceptional_line}(d) and (i) and Figs.~\ref{fig:exceptional_line}(e) and (j), the eigenvalues fit well to a square root function near the EPs, showing that the singularities in our system exhibit the characteristics of an EP2.


Both the real and imaginary parts of $E_{\pm}$ can be extracted by fitting the vacuum Rabi spectrum.
The cavity transmission spectrum, as a function of the probe laser frequency, is expressed as 
\begin{equation}
	T = \left| \frac{\kappa(\Delta_{\textrm{pc}} +i\gamma)}{(E_{+} - \Delta_{\textrm{pc}})(E_{-} - \Delta_\textrm{pc})} \right|^{2}.
	\label{eq:transmission}
\end{equation}

\noindent
Given values of $\kappa$  and $\gamma$, we numerically solve the master equation of the atom-cavity system using the quantum Monte Carlo (QMC) method. 
This yields the vacuum Rabi spectrum measured at the cavity transmission, which is then fitted to Eq.~\eqref{eq:transmission}, providing both the real and imaginary parts of $E_{+}$ and $E_{-}$.
The results are shown in Figs.~\ref{fig:exceptional_line}(a)--(c) and (f)--(h), demonstrating good agreement between the extracted eigenvalues and those of non-Hermitian Hamiltonian Eq.~\eqref{eq:H_nH}.
This finding confirms that, in practice, the eigenvalues of Eq.~\eqref{eq:H_nH} can be obtained from measurements and fittings of the cavity transmission spectrum.
We provide a detailed description of the derivation of Eq.~\eqref{eq:transmission} and the QMC simulation in Ref.~\cite{SI}.


Next, we examine the tunability of the singular point on the Riemann surface in the parameter space defined by $\Delta_{\textrm{ca}}(=\omega_{\textrm{c}} - \omega_{\textrm{a}})$ and $g$.
Fig.~\ref{fig:Riemann_surface}(a) shows the Riemann surface formed by the eigenvalues at $h=0$.
In this configuration, an EP appears at $g/(2\pi) = 4.99$~MHz and $\Delta_{\textrm{ca}}=0$~MHz. 
As the nanotip intrudes into the cavity mode, the enhanced $\kappa$ changes the EP condition according to the relation of $g=|\kappa-\gamma|/2$. 
As presented in Figs.~\ref{fig:Riemann_surface}(b) and (c), we observe that the EP position moves to $g/(2\pi) = 65.0$~MHz at $h=6.8$~$\mu$m, and further to $g/(2\pi) = 121.5$~MHz at $h=7.0$~$\mu$m.
These results demonstrate that the location of the EP can be adjusted by tuning the cavity dissipation rate.
Also, the results of the QMC simulations agree well with the eigenvalues of Eq.~\eqref{eq:H_nH}, showing that such behaviors can be experimentally observed by measuring and fitting the vacuum Rabi spectrum. 

\begin{figure*}[!t]
	\includegraphics[width=1\linewidth]{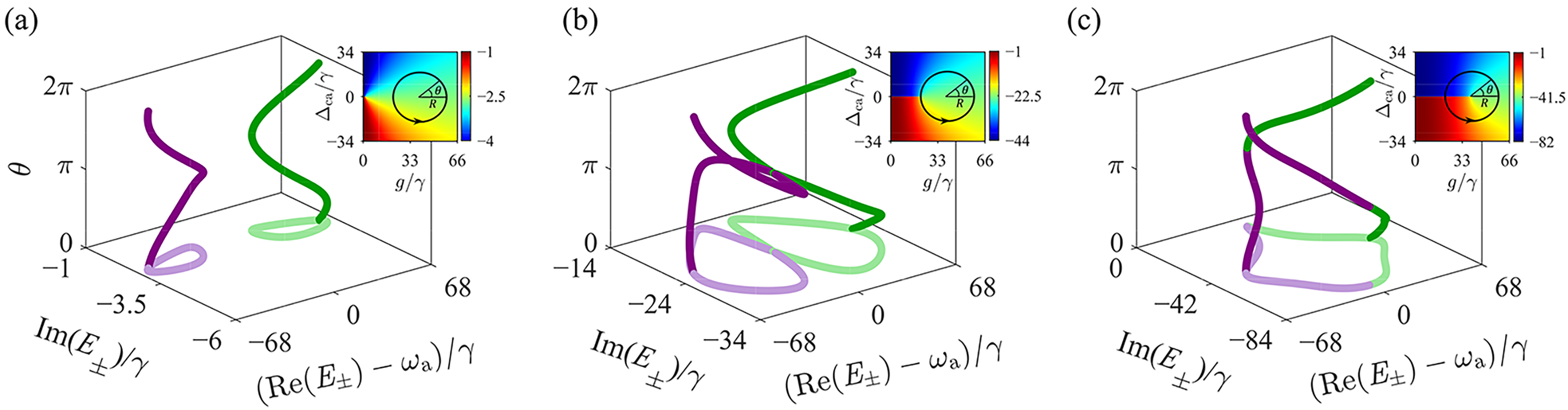}
	\caption{
		(a)--(c) Braid plots illustrating both the real and imaginary parts of $E_{+}$ and $E_{-}$ as $\theta$ changes, at $h=5$~$\mu$m ($\kappa/(2\pi) = 12.7$~MHz), $6.8$~$\mu$m ($133$~MHz), and $7.0$~$\mu$m ($246$~MHz), respectively. 
		Green and purple curves denote $E_{+}$ and $E_{-}$, respectively. 			
		The projections of braids onto the Re($E_{\pm}$) and Im($E_{\pm}$) plane are shown in light green and purple curves.
		(inset) Riemann surface seen from top. 
		Color code indicates Im($E_{\pm})/\gamma$, and black line presents the evolution of $\theta$ at $R/(2\pi) = 56.5$~MHz.
		EP position corresponds to $(g, \Delta_{\textrm{ca}})/(2\pi) = (5.0, 0)$, $(65.0, 0)$, and $(121.5, 0)$~MHz, respectively.
		The center of the circle is $(g, \Delta_{\textrm{ca}})/(2\pi) = (121.5, 0)$~MHz.
		}
	\label{fig:braid}			
\end{figure*}


In order to investigate topological properties on the Riemann surfaces, we vary the system parameters along a circular loop defined as $g = g_{\textrm{c}} + R\cos(\theta)$ and $\Delta_{\textrm{ca}}  = \Delta_0 + R\sin(\theta)$, where $g_\textrm{c}/(2\pi) = 121.5$, $\Delta_0 / (2\pi) = 0$, and $R / (2\pi) = 56.5$~MHz.  
If the loop does not intersect an EP or a branch cut, as shown in Fig.~\ref{fig:braid}(a), the eigenvalues form a braid in the space of $\theta$, $g$, and $\Delta_{\textrm{ca}}$.  
The projection of the braid onto the $\textrm{Re}\left(E_{\pm}\right)$-$\textrm{Im}\left(E_{\pm}\right)$ plane reveals a gap between the eigenvalues, resulting in two separate unknots.

We now introduce a topological invariant, the winding number~\cite{SI, Ding2022}, which characterizes the phase of the eigenvalues  
\begin{equation}
	\mathcal{W} = \frac{1}{2\pi i} \oint_{C_z} \frac{\partial \log (E_{+}(z) + E_{-}(z) )}{\partial z}\, dz,
	\label{eq:winding_number}
\end{equation}

\noindent 
where $z = \Delta_{\textrm{ca}}/2 + ig$, and $C_z$ is a closed contour in the complex plane.  
In the case of Fig.~\ref{fig:braid}(a), where the loop neither encloses an EP nor intersects an EP or branch cut, we find $\mathcal{W} = 0$, indicating a topologically trivial phase.

When the nanotip is further inserted to a depth of $h=6.8$~$\mu$m, the shifted position of the EP lies on the closed loop, as depicted in Fig.~\ref{fig:braid}(b).
At this point, $\mathcal{W}$ becomes ill-defined, marking the transition boundary between the topologically trivial and nontrivial phases.

At the nanotip position of $h=7$~$\mu$m (Fig.~\ref{fig:braid}(c)), the EP is within the closed parametric loop. 
Two notable features are pointed out in this configuration. 
First, as $\theta$ increases from $0$, the trajectory crosses the branch cut at $\theta = \pi$. 
At this point, while the real parts of the eigenvalues evolve continuously, the imaginary parts undergo a swapping between $|+\rangle$ and $|-\rangle$ states. 
This interchange is evident in Fig.~\ref{fig:braid}(c), where the green and purple curves exchange their positions. 
Second, because the loop encircles the EP, the topological winding number $\mathcal{W}=+1$, signifying a nontrivial topological phase. 
Consequently, a quantum state can acquire a geometric phase of $\pi$ through this closed evolution.
The detailed calculation of $\mathcal{W}$ is provided in Ref.~\cite{SI}.


We remark the following points in relation to our study.
First, differently from the works regarding inserting nanoscopic~\cite{Jayich2008, Zhang2018a} or macroscopic structures~\cite{Xu2024} into optical cavities, we focus on the possible application of EP in the quantum regime.
In addition to Refs.~\cite{Silveri2019, Partanen2019, Maurya2024}, our method can be harnessed to explore quantum state tomography~\cite{Naghiloo2019}, mode switching~\cite{Liu2021}, quantum jumps~\cite{Chen2021}, and other various applications~\cite{Lu2025} near EP.
Second, for the choice of the gold nanotip, metallic materials are preferred over dielectrics due to their significantly higher reflectivity, allowing even a small displacement of the tip to produce a noticeable change in $\kappa$.
Among metals, gold is selected with feasible ion-cavity experiments in mind~\cite{Lee2019a, Ong2020, Takahashi2020}.
Gold is particularly advantageous due to its minimal oxidation, which helps reduce the influence of stray charges on the ion.
The tip width of $300$~nm is chosen to ensure practical control of the tip’s position.
For instance, when $\kappa$ varies substantially between $h = 6.4$ and $7.0$~$\mu$m, the nanotip position can be reliably controlled at this scale using a piezoelectric stage.
Third, from the perspective of fundamental quantum optics, introducing a controllable perturber provides an additional degree of freedom for tuning the Purcell factor, which is proportional to the cavity's quality factor.
Also, it was proposed that, an electromagnetic wave can be virtually perfectly absorbed by modulating the cavity decay rate~\cite{Sounas2020}.
Lastly, as shown in Fig.~\ref{fig:setup}(a), Figs.~\ref{fig:field_calculation}(g) and (h), and Ref.~\cite{SI}, the scattered light can be measured by, e.~g., placing lenses and a photodetector along the upward transverse direction of the cavity. 
Such setup provides an additional channel for detecting cavity photons, complementing the conventional cavity transmission measurement.
This controls the extraction rate of the information from an atomic quantum memory to photons.
Furthermore, since it offers an additional degree of freedom for entanglement generation, if one can induce polarization-dependent perturber, our approach opens the possibility for realizing a high-dimensional entanglement, among an atom, photonic polarization, and spatial mode.
In summary, we have theoretically investigated an experimental method for tuning the decay rate of an atom-cavity system, by introducing a gold nanotip into the resonator mode.
Our analysis reveals that the cavity dissipation rate can be enhanced by approximately a factor of $20$, owing to the scattering loss caused by the tip.
Moreover, by varying the dissipation rate, the position of the EP can be adjusted, giving rise to a topological transition driven solely by dissipation control.
Beyond these results, we anticipate that our approach can be utilized for exploring a variety of non-Hermitian quantum phenomena near EPs, with single atoms or ions coupled to an optical resonator.

\begin{acknowledgments}
This work has been supported by BK21 FOUR program, National Research Foundation (Grant Nos.~RS-2019-NR040049, RS-2023-00291684, and RS-2024-00442855), the Institute for Information \& Communications Technology Planning \&	Evaluation (IITP, Grant Nos.~RS-2022-II221040 and RS-2025-02307012), and Samsung Electronics Co., Ltd.~(Grant No.~IO201211-08121-01).
\end{acknowledgments}

\section*{Data Availability}
Our data are available at https://doi.org/10.5281/zenodo.15622402.

\section*{References}

\bibliographystyle{apsrev4-2}
\bibliography{bibliography}

\begin{thebibliography}{39}%
\makeatletter
\providecommand \@ifxundefined [1]{%
 \@ifx{#1\undefined}
}%
\providecommand \@ifnum [1]{%
 \ifnum #1\expandafter \@firstoftwo
 \else \expandafter \@secondoftwo
 \fi
}%
\providecommand \@ifx [1]{%
 \ifx #1\expandafter \@firstoftwo
 \else \expandafter \@secondoftwo
 \fi
}%
\providecommand \natexlab [1]{#1}%
\providecommand \enquote  [1]{``#1''}%
\providecommand \bibnamefont  [1]{#1}%
\providecommand \bibfnamefont [1]{#1}%
\providecommand \citenamefont [1]{#1}%
\providecommand \href@noop [0]{\@secondoftwo}%
\providecommand \href [0]{\begingroup \@sanitize@url \@href}%
\providecommand \@href[1]{\@@startlink{#1}\@@href}%
\providecommand \@@href[1]{\endgroup#1\@@endlink}%
\providecommand \@sanitize@url [0]{\catcode `\\12\catcode `\$12\catcode
  `\&12\catcode `\#12\catcode `\^12\catcode `\_12\catcode `\%12\relax}%
\providecommand \@@startlink[1]{}%
\providecommand \@@endlink[0]{}%
\providecommand \url  [0]{\begingroup\@sanitize@url \@url }%
\providecommand \@url [1]{\endgroup\@href {#1}{\urlprefix }}%
\providecommand \urlprefix  [0]{URL }%
\providecommand \Eprint [0]{\href }%
\providecommand \doibase [0]{https://doi.org/}%
\providecommand \selectlanguage [0]{\@gobble}%
\providecommand \bibinfo  [0]{\@secondoftwo}%
\providecommand \bibfield  [0]{\@secondoftwo}%
\providecommand \translation [1]{[#1]}%
\providecommand \BibitemOpen [0]{}%
\providecommand \bibitemStop [0]{}%
\providecommand \bibitemNoStop [0]{.\EOS\space}%
\providecommand \EOS [0]{\spacefactor3000\relax}%
\providecommand \BibitemShut  [1]{\csname bibitem#1\endcsname}%
\let\auto@bib@innerbib\@empty
\bibitem [{\citenamefont {Gardiner}\ and\ \citenamefont
  {Zoller}(2004)}]{Gardiner04}%
  \BibitemOpen
  \bibfield  {author} {\bibinfo {author} {\bibfnamefont {C.}~\bibnamefont
  {Gardiner}}\ and\ \bibinfo {author} {\bibfnamefont {P.}~\bibnamefont
  {Zoller}},\ }\href {http://books.google.at/books?id=a\_xsT8oGhdgC} {\emph
  {\bibinfo {title} {Quantum Noise: A Handbook of Markovian and Non-Markovian
  Quantum Stochastic Methods with Applications to Quantum Optics}}},\ Springer
  Series in Synergetics\ (\bibinfo  {publisher} {Springer-Verlag Berlin
  Heidelberg},\ \bibinfo {address} {Berlin},\ \bibinfo {year}
  {2004})\BibitemShut {NoStop}%
\bibitem [{\citenamefont {Plenio}\ and\ \citenamefont
  {Knight}(1998)}]{Plenio1998}%
  \BibitemOpen
  \bibfield  {author} {\bibinfo {author} {\bibfnamefont {M.~B.}\ \bibnamefont
  {Plenio}}\ and\ \bibinfo {author} {\bibfnamefont {P.~L.}\ \bibnamefont
  {Knight}},\ }\href {https://doi.org/10.1103/RevModPhys.70.101} {\bibfield
  {journal} {\bibinfo  {journal} {Rev. Mod. Phys.}\ }\textbf {\bibinfo {volume}
  {70}},\ \bibinfo {pages} {101} (\bibinfo {year} {1998})}\BibitemShut
  {NoStop}%
\bibitem [{\citenamefont {Carmichael}(2008)}]{Carmichael2008}%
  \BibitemOpen
  \bibfield  {author} {\bibinfo {author} {\bibfnamefont {H.~J.}\ \bibnamefont
  {Carmichael}},\ }\href {https://bib-pubdb1.desy.de/record/365946} {\emph
  {\bibinfo {title} {{S}tatistical methods in quantum optics: {V}ol. 2:
  {N}on-classical fields}}},\ Theoretical and mathematical physics\ (\bibinfo
  {publisher} {Springer},\ \bibinfo {address} {Berlin},\ \bibinfo {year}
  {2008})\BibitemShut {NoStop}%
\bibitem [{\citenamefont {Kimble}(2008)}]{Kimble08a}%
  \BibitemOpen
  \bibfield  {author} {\bibinfo {author} {\bibfnamefont {H.~J.}\ \bibnamefont
  {Kimble}},\ }\href@noop {} {\bibfield  {journal} {\bibinfo  {journal}
  {Nature}\ }\textbf {\bibinfo {volume} {453}},\ \bibinfo {pages} {1023}
  (\bibinfo {year} {2008})}\BibitemShut {NoStop}%
\bibitem [{\citenamefont {Reiserer}\ and\ \citenamefont
  {Rempe}(2015)}]{Reiserer2015}%
  \BibitemOpen
  \bibfield  {author} {\bibinfo {author} {\bibfnamefont {A.}~\bibnamefont
  {Reiserer}}\ and\ \bibinfo {author} {\bibfnamefont {G.}~\bibnamefont
  {Rempe}},\ }\href {https://doi.org/10.1103/RevModPhys.87.1379} {\bibfield
  {journal} {\bibinfo  {journal} {Rev. Mod. Phys.}\ }\textbf {\bibinfo {volume}
  {87}},\ \bibinfo {pages} {1379} (\bibinfo {year} {2015})}\BibitemShut
  {NoStop}%
\bibitem [{\citenamefont {Plenio}\ \emph {et~al.}(1999)\citenamefont {Plenio},
  \citenamefont {Huelga}, \citenamefont {Beige},\ and\ \citenamefont
  {Knight}}]{Plenio99}%
  \BibitemOpen
  \bibfield  {author} {\bibinfo {author} {\bibfnamefont {M.~B.}\ \bibnamefont
  {Plenio}}, \bibinfo {author} {\bibfnamefont {S.~F.}\ \bibnamefont {Huelga}},
  \bibinfo {author} {\bibfnamefont {A.}~\bibnamefont {Beige}},\ and\ \bibinfo
  {author} {\bibfnamefont {P.~L.}\ \bibnamefont {Knight}},\ }\href
  {https://doi.org/10.1103/PhysRevA.59.2468} {\bibfield  {journal} {\bibinfo
  {journal} {Phys. Rev. A}\ }\textbf {\bibinfo {volume} {59}},\ \bibinfo
  {pages} {2468} (\bibinfo {year} {1999})}\BibitemShut {NoStop}%
\bibitem [{\citenamefont {Kastoryano}\ \emph {et~al.}(2011)\citenamefont
  {Kastoryano}, \citenamefont {Reiter},\ and\ \citenamefont
  {S\o{}rensen}}]{Kastoryano11}%
  \BibitemOpen
  \bibfield  {author} {\bibinfo {author} {\bibfnamefont {M.~J.}\ \bibnamefont
  {Kastoryano}}, \bibinfo {author} {\bibfnamefont {F.}~\bibnamefont {Reiter}},\
  and\ \bibinfo {author} {\bibfnamefont {A.~S.}\ \bibnamefont {S\o{}rensen}},\
  }\href {https://doi.org/10.1103/PhysRevLett.106.090502} {\bibfield  {journal}
  {\bibinfo  {journal} {Phys. Rev. Lett.}\ }\textbf {\bibinfo {volume} {106}},\
  \bibinfo {pages} {090502} (\bibinfo {year} {2011})}\BibitemShut {NoStop}%
\bibitem [{\citenamefont {Su}\ \emph {et~al.}(2014)\citenamefont {Su},
  \citenamefont {Shao}, \citenamefont {Wang},\ and\ \citenamefont
  {Zhang}}]{Su2014}%
  \BibitemOpen
  \bibfield  {author} {\bibinfo {author} {\bibfnamefont {S.-L.}\ \bibnamefont
  {Su}}, \bibinfo {author} {\bibfnamefont {X.-Q.}\ \bibnamefont {Shao}},
  \bibinfo {author} {\bibfnamefont {H.-F.}\ \bibnamefont {Wang}},\ and\
  \bibinfo {author} {\bibfnamefont {S.}~\bibnamefont {Zhang}},\ }\href
  {https://doi.org/10.1103/PhysRevA.90.054302} {\bibfield  {journal} {\bibinfo
  {journal} {Phys. Rev. A}\ }\textbf {\bibinfo {volume} {90}},\ \bibinfo
  {pages} {054302} (\bibinfo {year} {2014})}\BibitemShut {NoStop}%
\bibitem [{\citenamefont {Wolke}\ \emph {et~al.}(2012)\citenamefont {Wolke},
  \citenamefont {Klinner}, \citenamefont {Keßler},\ and\ \citenamefont
  {Hemmerich}}]{Wolke2012}%
  \BibitemOpen
  \bibfield  {author} {\bibinfo {author} {\bibfnamefont {M.}~\bibnamefont
  {Wolke}}, \bibinfo {author} {\bibfnamefont {J.}~\bibnamefont {Klinner}},
  \bibinfo {author} {\bibfnamefont {H.}~\bibnamefont {Keßler}},\ and\ \bibinfo
  {author} {\bibfnamefont {A.}~\bibnamefont {Hemmerich}},\ }\href
  {https://doi.org/10.1126/science.1219166} {\bibfield  {journal} {\bibinfo
  {journal} {Science}\ }\textbf {\bibinfo {volume} {337}},\ \bibinfo {pages}
  {75} (\bibinfo {year} {2012})}\BibitemShut {NoStop}%
\bibitem [{\citenamefont {Wellnitz}\ \emph {et~al.}(2020)\citenamefont
  {Wellnitz}, \citenamefont {Sch\"utz}, \citenamefont {Whitlock}, \citenamefont
  {Schachenmayer},\ and\ \citenamefont {Pupillo}}]{Wellnitz2020}%
  \BibitemOpen
  \bibfield  {author} {\bibinfo {author} {\bibfnamefont {D.}~\bibnamefont
  {Wellnitz}}, \bibinfo {author} {\bibfnamefont {S.}~\bibnamefont {Sch\"utz}},
  \bibinfo {author} {\bibfnamefont {S.}~\bibnamefont {Whitlock}}, \bibinfo
  {author} {\bibfnamefont {J.}~\bibnamefont {Schachenmayer}},\ and\ \bibinfo
  {author} {\bibfnamefont {G.}~\bibnamefont {Pupillo}},\ }\href
  {https://doi.org/10.1103/PhysRevLett.125.193201} {\bibfield  {journal}
  {\bibinfo  {journal} {Phys. Rev. Lett.}\ }\textbf {\bibinfo {volume} {125}},\
  \bibinfo {pages} {193201} (\bibinfo {year} {2020})}\BibitemShut {NoStop}%
\bibitem [{\citenamefont {Choi}\ \emph {et~al.}(2010)\citenamefont {Choi},
  \citenamefont {Kang}, \citenamefont {Lim}, \citenamefont {Kim}, \citenamefont
  {Kim}, \citenamefont {Lee},\ and\ \citenamefont {An}}]{Choi2010}%
  \BibitemOpen
  \bibfield  {author} {\bibinfo {author} {\bibfnamefont {Y.}~\bibnamefont
  {Choi}}, \bibinfo {author} {\bibfnamefont {S.}~\bibnamefont {Kang}}, \bibinfo
  {author} {\bibfnamefont {S.}~\bibnamefont {Lim}}, \bibinfo {author}
  {\bibfnamefont {W.}~\bibnamefont {Kim}}, \bibinfo {author} {\bibfnamefont
  {J.-R.}\ \bibnamefont {Kim}}, \bibinfo {author} {\bibfnamefont {J.-H.}\
  \bibnamefont {Lee}},\ and\ \bibinfo {author} {\bibfnamefont {K.}~\bibnamefont
  {An}},\ }\href {https://doi.org/10.1103/PhysRevLett.104.153601} {\bibfield
  {journal} {\bibinfo  {journal} {Phys. Rev. Lett.}\ }\textbf {\bibinfo
  {volume} {104}},\ \bibinfo {pages} {153601} (\bibinfo {year}
  {2010})}\BibitemShut {NoStop}%
\bibitem [{\citenamefont {Lu}\ \emph {et~al.}(2018)\citenamefont {Lu},
  \citenamefont {Peng}, \citenamefont {Cao}, \citenamefont {Xu}, \citenamefont
  {Wiersig}, \citenamefont {Gong},\ and\ \citenamefont {Xiao}}]{Lu2018}%
  \BibitemOpen
  \bibfield  {author} {\bibinfo {author} {\bibfnamefont {Y.-K.}\ \bibnamefont
  {Lu}}, \bibinfo {author} {\bibfnamefont {P.}~\bibnamefont {Peng}}, \bibinfo
  {author} {\bibfnamefont {Q.-T.}\ \bibnamefont {Cao}}, \bibinfo {author}
  {\bibfnamefont {D.}~\bibnamefont {Xu}}, \bibinfo {author} {\bibfnamefont
  {J.}~\bibnamefont {Wiersig}}, \bibinfo {author} {\bibfnamefont
  {Q.}~\bibnamefont {Gong}},\ and\ \bibinfo {author} {\bibfnamefont {Y.-F.}\
  \bibnamefont {Xiao}},\ }\href {https://doi.org/10.1016/j.scib.2018.07.020}
  {\bibfield  {journal} {\bibinfo  {journal} {Sci. Bull.}\ }\textbf {\bibinfo
  {volume} {63}},\ \bibinfo {pages} {1096} (\bibinfo {year}
  {2018})}\BibitemShut {NoStop}%
\bibitem [{\citenamefont {Huang}\ \emph {et~al.}(2022)\citenamefont {Huang},
  \citenamefont {\''Ozdemir}, \citenamefont {Liao}, \citenamefont {Minganti},
  \citenamefont {Kuang}, \citenamefont {Nori},\ and\ \citenamefont
  {Jing}}]{Huang2022}%
  \BibitemOpen
  \bibfield  {author} {\bibinfo {author} {\bibfnamefont {R.}~\bibnamefont
  {Huang}}, \bibinfo {author} {\bibfnamefont {S.~K.}\ \bibnamefont
  {\''Ozdemir}}, \bibinfo {author} {\bibfnamefont {J.-Q.}\ \bibnamefont
  {Liao}}, \bibinfo {author} {\bibfnamefont {F.}~\bibnamefont {Minganti}},
  \bibinfo {author} {\bibfnamefont {L.-M.}\ \bibnamefont {Kuang}}, \bibinfo
  {author} {\bibfnamefont {F.}~\bibnamefont {Nori}},\ and\ \bibinfo {author}
  {\bibfnamefont {H.}~\bibnamefont {Jing}},\ }\href
  {https://doi.org/https://doi.org/10.1002/lpor.202100430} {\bibfield
  {journal} {\bibinfo  {journal} {Laser Photonics Rev.}\ }\textbf {\bibinfo
  {volume} {16}},\ \bibinfo {pages} {2100430} (\bibinfo {year}
  {2022})}\BibitemShut {NoStop}%
\bibitem [{\citenamefont {Li}\ \emph {et~al.}(2023)\citenamefont {Li},
  \citenamefont {Li}, \citenamefont {Zhang},\ and\ \citenamefont
  {Zhong}}]{Li2023}%
  \BibitemOpen
  \bibfield  {author} {\bibinfo {author} {\bibfnamefont {Z.}~\bibnamefont
  {Li}}, \bibinfo {author} {\bibfnamefont {X.}~\bibnamefont {Li}}, \bibinfo
  {author} {\bibfnamefont {G.}~\bibnamefont {Zhang}},\ and\ \bibinfo {author}
  {\bibfnamefont {X.}~\bibnamefont {Zhong}},\ }\href
  {https://doi.org/10.3389/fphy.2023.1168372} {\bibfield  {journal} {\bibinfo
  {journal} {Front. Phys.}\ }\textbf {\bibinfo {volume} {11}},\ \bibinfo
  {pages} {1168372} (\bibinfo {year} {2023})}\BibitemShut {NoStop}%
\bibitem [{\citenamefont {Kim}\ \emph {et~al.}(2023)\citenamefont {Kim},
  \citenamefont {Ha}, \citenamefont {Kim}, \citenamefont {Lee}, \citenamefont
  {Lee}, \citenamefont {Won}, \citenamefont {Moon},\ and\ \citenamefont
  {Lee}}]{Kim2023a}%
  \BibitemOpen
  \bibfield  {author} {\bibinfo {author} {\bibfnamefont {J.}~\bibnamefont
  {Kim}}, \bibinfo {author} {\bibfnamefont {T.}~\bibnamefont {Ha}}, \bibinfo
  {author} {\bibfnamefont {D.}~\bibnamefont {Kim}}, \bibinfo {author}
  {\bibfnamefont {D.}~\bibnamefont {Lee}}, \bibinfo {author} {\bibfnamefont
  {K.-S.}\ \bibnamefont {Lee}}, \bibinfo {author} {\bibfnamefont
  {J.}~\bibnamefont {Won}}, \bibinfo {author} {\bibfnamefont {Y.}~\bibnamefont
  {Moon}},\ and\ \bibinfo {author} {\bibfnamefont {M.}~\bibnamefont {Lee}},\
  }\href {https://doi.org/10.1063/5.0168372} {\bibfield  {journal} {\bibinfo
  {journal} {Appl. Phys. Lett.}\ }\textbf {\bibinfo {volume} {123}},\ \bibinfo
  {pages} {161104} (\bibinfo {year} {2023})}\BibitemShut {NoStop}%
\bibitem [{\citenamefont {Lee}\ \emph {et~al.}(2024)\citenamefont {Lee},
  \citenamefont {Kim},\ and\ \citenamefont {An}}]{Lee2024}%
  \BibitemOpen
  \bibfield  {author} {\bibinfo {author} {\bibfnamefont {J.}~\bibnamefont
  {Lee}}, \bibinfo {author} {\bibfnamefont {J.}~\bibnamefont {Kim}},\ and\
  \bibinfo {author} {\bibfnamefont {K.}~\bibnamefont {An}},\ }\href
  {https://doi.org/10.1038/s41598-024-54008-w} {\bibfield  {journal} {\bibinfo
  {journal} {Sci. Rep.}\ }\textbf {\bibinfo {volume} {14}},\ \bibinfo {pages}
  {3471} (\bibinfo {year} {2024})}\BibitemShut {NoStop}%
\bibitem [{\citenamefont {Agarwal}(2024)}]{Agarwal2024}%
  \BibitemOpen
  \bibfield  {author} {\bibinfo {author} {\bibfnamefont {G.~S.}\ \bibnamefont
  {Agarwal}},\ }\href {https://doi.org/10.1103/PhysRevResearch.6.L012050}
  {\bibfield  {journal} {\bibinfo  {journal} {Phys. Rev. Res.}\ }\textbf
  {\bibinfo {volume} {6}},\ \bibinfo {pages} {L012050} (\bibinfo {year}
  {2024})}\BibitemShut {NoStop}%
\bibitem [{\citenamefont {Ferri}\ \emph {et~al.}(2021)\citenamefont {Ferri},
  \citenamefont {Rosa-Medina}, \citenamefont {Finger}, \citenamefont {Dogra},
  \citenamefont {Soriente}, \citenamefont {Zilberberg}, \citenamefont
  {Donner},\ and\ \citenamefont {Esslinger}}]{Ferri2021}%
  \BibitemOpen
  \bibfield  {author} {\bibinfo {author} {\bibfnamefont {F.}~\bibnamefont
  {Ferri}}, \bibinfo {author} {\bibfnamefont {R.}~\bibnamefont {Rosa-Medina}},
  \bibinfo {author} {\bibfnamefont {F.}~\bibnamefont {Finger}}, \bibinfo
  {author} {\bibfnamefont {N.}~\bibnamefont {Dogra}}, \bibinfo {author}
  {\bibfnamefont {M.}~\bibnamefont {Soriente}}, \bibinfo {author}
  {\bibfnamefont {O.}~\bibnamefont {Zilberberg}}, \bibinfo {author}
  {\bibfnamefont {T.}~\bibnamefont {Donner}},\ and\ \bibinfo {author}
  {\bibfnamefont {T.}~\bibnamefont {Esslinger}},\ }\href
  {https://doi.org/10.1103/PhysRevX.11.041046} {\bibfield  {journal} {\bibinfo
  {journal} {Phys. Rev. X}\ }\textbf {\bibinfo {volume} {11}},\ \bibinfo
  {pages} {041046} (\bibinfo {year} {2021})}\BibitemShut {NoStop}%
\bibitem [{\citenamefont {So}\ \emph {et~al.}(2024)\citenamefont {So},
  \citenamefont {Suganthi}, \citenamefont {Menon}, \citenamefont {Zhu},
  \citenamefont {Zhuravel}, \citenamefont {Pu}, \citenamefont {Wolynes},
  \citenamefont {Onuchic},\ and\ \citenamefont {Pagano}}]{So2024}%
  \BibitemOpen
  \bibfield  {author} {\bibinfo {author} {\bibfnamefont {V.}~\bibnamefont
  {So}}, \bibinfo {author} {\bibfnamefont {M.~D.}\ \bibnamefont {Suganthi}},
  \bibinfo {author} {\bibfnamefont {A.}~\bibnamefont {Menon}}, \bibinfo
  {author} {\bibfnamefont {M.}~\bibnamefont {Zhu}}, \bibinfo {author}
  {\bibfnamefont {R.}~\bibnamefont {Zhuravel}}, \bibinfo {author}
  {\bibfnamefont {H.}~\bibnamefont {Pu}}, \bibinfo {author} {\bibfnamefont
  {P.~G.}\ \bibnamefont {Wolynes}}, \bibinfo {author} {\bibfnamefont {J.~N.}\
  \bibnamefont {Onuchic}},\ and\ \bibinfo {author} {\bibfnamefont
  {G.}~\bibnamefont {Pagano}},\ }\href {https://doi.org/10.1126/sciadv.ads8011}
  {\bibfield  {journal} {\bibinfo  {journal} {Sci. Adv.}\ }\textbf {\bibinfo
  {volume} {10}},\ \bibinfo {pages} {eads8011} (\bibinfo {year}
  {2024})}\BibitemShut {NoStop}%
\bibitem [{\citenamefont {Li}\ \emph {et~al.}(2024)\citenamefont {Li},
  \citenamefont {Wu}, \citenamefont {Zhou}, \citenamefont {Zhang},
  \citenamefont {Zhao}, \citenamefont {Yuan}, \citenamefont {Cheng},
  \citenamefont {Li}, \citenamefont {Qin}, \citenamefont {Rong}, \citenamefont
  {Lin},\ and\ \citenamefont {Du}}]{Li2024}%
  \BibitemOpen
  \bibfield  {author} {\bibinfo {author} {\bibfnamefont {Y.}~\bibnamefont
  {Li}}, \bibinfo {author} {\bibfnamefont {Y.}~\bibnamefont {Wu}}, \bibinfo
  {author} {\bibfnamefont {Y.}~\bibnamefont {Zhou}}, \bibinfo {author}
  {\bibfnamefont {M.}~\bibnamefont {Zhang}}, \bibinfo {author} {\bibfnamefont
  {X.}~\bibnamefont {Zhao}}, \bibinfo {author} {\bibfnamefont {Y.}~\bibnamefont
  {Yuan}}, \bibinfo {author} {\bibfnamefont {X.}~\bibnamefont {Cheng}},
  \bibinfo {author} {\bibfnamefont {Y.}~\bibnamefont {Li}}, \bibinfo {author}
  {\bibfnamefont {X.}~\bibnamefont {Qin}}, \bibinfo {author} {\bibfnamefont
  {X.}~\bibnamefont {Rong}}, \bibinfo {author} {\bibfnamefont {Y.}~\bibnamefont
  {Lin}},\ and\ \bibinfo {author} {\bibfnamefont {J.}~\bibnamefont {Du}},\
  }\href@noop {} {\bibfield  {journal} {\bibinfo  {journal} {arXiv:2412.09776}\
  } (\bibinfo {year} {2024})}\BibitemShut {NoStop}%
\bibitem [{\citenamefont {Jones}\ \emph {et~al.}(2013)\citenamefont {Jones},
  \citenamefont {Huhtam\''aki}, \citenamefont {Salmilehto}, \citenamefont
  {Tan},\ and\ \citenamefont {M\''ott\''onen}}]{Jones2013}%
  \BibitemOpen
  \bibfield  {author} {\bibinfo {author} {\bibfnamefont {P.~J.}\ \bibnamefont
  {Jones}}, \bibinfo {author} {\bibfnamefont {J.~A.~M.}\ \bibnamefont
  {Huhtam\''aki}}, \bibinfo {author} {\bibfnamefont {J.}~\bibnamefont
  {Salmilehto}}, \bibinfo {author} {\bibfnamefont {K.~Y.}\ \bibnamefont
  {Tan}},\ and\ \bibinfo {author} {\bibfnamefont {M.}~\bibnamefont
  {M\''ott\''onen}},\ }\href {https://doi.org/10.1038/srep01987} {\bibfield
  {journal} {\bibinfo  {journal} {Sci. Rep.}\ }\textbf {\bibinfo {volume}
  {3}},\ \bibinfo {pages} {1987} (\bibinfo {year} {2013})}\BibitemShut
  {NoStop}%
\bibitem [{\citenamefont {Silveri}\ \emph {et~al.}(2019)\citenamefont
  {Silveri}, \citenamefont {Masuda}, \citenamefont {Sevriuk}, \citenamefont
  {Tan}, \citenamefont {Jenei}, \citenamefont {Hyyppä}, \citenamefont
  {Hassler}, \citenamefont {Partanen}, \citenamefont {Goetz}, \citenamefont
  {Lake}, \citenamefont {Grönberg},\ and\ \citenamefont
  {Möttönen}}]{Silveri2019}%
  \BibitemOpen
  \bibfield  {author} {\bibinfo {author} {\bibfnamefont {M.}~\bibnamefont
  {Silveri}}, \bibinfo {author} {\bibfnamefont {S.}~\bibnamefont {Masuda}},
  \bibinfo {author} {\bibfnamefont {V.}~\bibnamefont {Sevriuk}}, \bibinfo
  {author} {\bibfnamefont {K.~Y.}\ \bibnamefont {Tan}}, \bibinfo {author}
  {\bibfnamefont {M.}~\bibnamefont {Jenei}}, \bibinfo {author} {\bibfnamefont
  {E.}~\bibnamefont {Hyyppä}}, \bibinfo {author} {\bibfnamefont
  {F.}~\bibnamefont {Hassler}}, \bibinfo {author} {\bibfnamefont
  {M.}~\bibnamefont {Partanen}}, \bibinfo {author} {\bibfnamefont
  {J.}~\bibnamefont {Goetz}}, \bibinfo {author} {\bibfnamefont {R.~E.}\
  \bibnamefont {Lake}}, \bibinfo {author} {\bibfnamefont {L.}~\bibnamefont
  {Grönberg}},\ and\ \bibinfo {author} {\bibfnamefont {M.}~\bibnamefont
  {Möttönen}},\ }\href {https://doi.org/10.1038/s41567-019-0449-0} {\bibfield
   {journal} {\bibinfo  {journal} {Nat. Phys.}\ }\textbf {\bibinfo {volume}
  {15}},\ \bibinfo {pages} {533} (\bibinfo {year} {2019})}\BibitemShut
  {NoStop}%
\bibitem [{\citenamefont {Partanen}\ \emph {et~al.}(2019)\citenamefont
  {Partanen}, \citenamefont {Goetz}, \citenamefont {Tan}, \citenamefont
  {Kohvakka}, \citenamefont {Sevriuk}, \citenamefont {Lake}, \citenamefont
  {Kokkoniemi}, \citenamefont {Ikonen}, \citenamefont {Hazra}, \citenamefont
  {M\"akinen}, \citenamefont {Hyypp\"a}, \citenamefont {Gr\"onberg},
  \citenamefont {Vesterinen}, \citenamefont {Silveri},\ and\ \citenamefont
  {M\"ott\"onen}}]{Partanen2019}%
  \BibitemOpen
  \bibfield  {author} {\bibinfo {author} {\bibfnamefont {M.}~\bibnamefont
  {Partanen}}, \bibinfo {author} {\bibfnamefont {J.}~\bibnamefont {Goetz}},
  \bibinfo {author} {\bibfnamefont {K.~Y.}\ \bibnamefont {Tan}}, \bibinfo
  {author} {\bibfnamefont {K.}~\bibnamefont {Kohvakka}}, \bibinfo {author}
  {\bibfnamefont {V.}~\bibnamefont {Sevriuk}}, \bibinfo {author} {\bibfnamefont
  {R.~E.}\ \bibnamefont {Lake}}, \bibinfo {author} {\bibfnamefont
  {R.}~\bibnamefont {Kokkoniemi}}, \bibinfo {author} {\bibfnamefont
  {J.}~\bibnamefont {Ikonen}}, \bibinfo {author} {\bibfnamefont
  {D.}~\bibnamefont {Hazra}}, \bibinfo {author} {\bibfnamefont
  {A.}~\bibnamefont {M\"akinen}}, \bibinfo {author} {\bibfnamefont
  {E.}~\bibnamefont {Hyypp\"a}}, \bibinfo {author} {\bibfnamefont
  {L.}~\bibnamefont {Gr\"onberg}}, \bibinfo {author} {\bibfnamefont
  {V.}~\bibnamefont {Vesterinen}}, \bibinfo {author} {\bibfnamefont
  {M.}~\bibnamefont {Silveri}},\ and\ \bibinfo {author} {\bibfnamefont
  {M.}~\bibnamefont {M\"ott\"onen}},\ }\href
  {https://doi.org/10.1103/PhysRevB.100.134505} {\bibfield  {journal} {\bibinfo
   {journal} {Phys. Rev. B}\ }\textbf {\bibinfo {volume} {100}},\ \bibinfo
  {pages} {134505} (\bibinfo {year} {2019})}\BibitemShut {NoStop}%
\bibitem [{\citenamefont {Maurya}\ \emph {et~al.}(2024)\citenamefont {Maurya},
  \citenamefont {Zhang}, \citenamefont {Kowsari}, \citenamefont {Kuo},
  \citenamefont {Hartsell}, \citenamefont {Miyamoto}, \citenamefont {Liu},
  \citenamefont {Shanto}, \citenamefont {Vlachos}, \citenamefont {Zarassi},
  \citenamefont {Murch},\ and\ \citenamefont {Levenson-Falk}}]{Maurya2024}%
  \BibitemOpen
  \bibfield  {author} {\bibinfo {author} {\bibfnamefont {V.}~\bibnamefont
  {Maurya}}, \bibinfo {author} {\bibfnamefont {H.}~\bibnamefont {Zhang}},
  \bibinfo {author} {\bibfnamefont {D.}~\bibnamefont {Kowsari}}, \bibinfo
  {author} {\bibfnamefont {A.}~\bibnamefont {Kuo}}, \bibinfo {author}
  {\bibfnamefont {D.~M.}\ \bibnamefont {Hartsell}}, \bibinfo {author}
  {\bibfnamefont {C.}~\bibnamefont {Miyamoto}}, \bibinfo {author}
  {\bibfnamefont {J.}~\bibnamefont {Liu}}, \bibinfo {author} {\bibfnamefont
  {S.}~\bibnamefont {Shanto}}, \bibinfo {author} {\bibfnamefont
  {E.}~\bibnamefont {Vlachos}}, \bibinfo {author} {\bibfnamefont
  {A.}~\bibnamefont {Zarassi}}, \bibinfo {author} {\bibfnamefont {K.~W.}\
  \bibnamefont {Murch}},\ and\ \bibinfo {author} {\bibfnamefont {E.~M.}\
  \bibnamefont {Levenson-Falk}},\ }\href
  {https://doi.org/10.1103/PRXQuantum.5.020321} {\bibfield  {journal} {\bibinfo
   {journal} {PRX Quantum}\ }\textbf {\bibinfo {volume} {5}},\ \bibinfo {pages}
  {020321} (\bibinfo {year} {2024})}\BibitemShut {NoStop}%
\bibitem [{\citenamefont {Lee}\ \emph {et~al.}(2014)\citenamefont {Lee},
  \citenamefont {Kim}, \citenamefont {Seo}, \citenamefont {Hong}, \citenamefont
  {Song}, \citenamefont {Dasari},\ and\ \citenamefont {An}}]{Lee2014}%
  \BibitemOpen
  \bibfield  {author} {\bibinfo {author} {\bibfnamefont {M.}~\bibnamefont
  {Lee}}, \bibinfo {author} {\bibfnamefont {J.}~\bibnamefont {Kim}}, \bibinfo
  {author} {\bibfnamefont {W.}~\bibnamefont {Seo}}, \bibinfo {author}
  {\bibfnamefont {H.-G.}\ \bibnamefont {Hong}}, \bibinfo {author}
  {\bibfnamefont {Y.}~\bibnamefont {Song}}, \bibinfo {author} {\bibfnamefont
  {R.~R.}\ \bibnamefont {Dasari}},\ and\ \bibinfo {author} {\bibfnamefont
  {K.}~\bibnamefont {An}},\ }\href {https://doi.org/10.1038/ncomms4441}
  {\bibfield  {journal} {\bibinfo  {journal} {Nat. Commun.}\ }\textbf {\bibinfo
  {volume} {5}},\ \bibinfo {pages} {3441} (\bibinfo {year} {2014})}\BibitemShut
  {NoStop}%
\bibitem [{\citenamefont {Guth\"{o}hrlein}\ \emph {et~al.}(2001)\citenamefont
  {Guth\"{o}hrlein}, \citenamefont {Keller}, \citenamefont {Hayasaka},
  \citenamefont {Lange},\ and\ \citenamefont {Walther}}]{Guthoehrlein01}%
  \BibitemOpen
  \bibfield  {author} {\bibinfo {author} {\bibfnamefont {G.~R.}\ \bibnamefont
  {Guth\"{o}hrlein}}, \bibinfo {author} {\bibfnamefont {M.}~\bibnamefont
  {Keller}}, \bibinfo {author} {\bibfnamefont {K.}~\bibnamefont {Hayasaka}},
  \bibinfo {author} {\bibfnamefont {W.}~\bibnamefont {Lange}},\ and\ \bibinfo
  {author} {\bibfnamefont {H.}~\bibnamefont {Walther}},\ }\href
  {https://doi.org/10.1038/35102129} {\bibfield  {journal} {\bibinfo  {journal}
  {Nature}\ }\textbf {\bibinfo {volume} {414}},\ \bibinfo {pages} {49}
  (\bibinfo {year} {2001})}\BibitemShut {NoStop}%
\bibitem [{SI()}]{SI}%
  \BibitemOpen
  \href@noop {} {\bibinfo  {journal} {See Supplemenatal Material}\
  }\BibitemShut {NoStop}%
\bibitem [{\citenamefont {Ding}\ \emph {et~al.}(2022)\citenamefont {Ding},
  \citenamefont {Fang},\ and\ \citenamefont {Ma}}]{Ding2022}%
  \BibitemOpen
\bibfield  {journal} {  }\bibfield  {author} {\bibinfo {author} {\bibfnamefont
  {K.}~\bibnamefont {Ding}}, \bibinfo {author} {\bibfnamefont {C.}~\bibnamefont
  {Fang}},\ and\ \bibinfo {author} {\bibfnamefont {G.}~\bibnamefont {Ma}},\
  }\href {https://doi.org/10.1038/s42254-022-00516-5} {\bibfield  {journal}
  {\bibinfo  {journal} {Nat. Rev. Phys.}\ }\textbf {\bibinfo {volume} {4}},\
  \bibinfo {pages} {745} (\bibinfo {year} {2022})}\BibitemShut {NoStop}%
\bibitem [{\citenamefont {Jayich}\ \emph {et~al.}(2008)\citenamefont {Jayich},
  \citenamefont {Sankey}, \citenamefont {Zwickl}, \citenamefont {Yang},
  \citenamefont {Thompson}, \citenamefont {Girvin}, \citenamefont {Clerk},
  \citenamefont {Marquardt},\ and\ \citenamefont {Harris}}]{Jayich2008}%
  \BibitemOpen
  \bibfield  {author} {\bibinfo {author} {\bibfnamefont {A.~M.}\ \bibnamefont
  {Jayich}}, \bibinfo {author} {\bibfnamefont {J.~C.}\ \bibnamefont {Sankey}},
  \bibinfo {author} {\bibfnamefont {B.~M.}\ \bibnamefont {Zwickl}}, \bibinfo
  {author} {\bibfnamefont {C.}~\bibnamefont {Yang}}, \bibinfo {author}
  {\bibfnamefont {J.~D.}\ \bibnamefont {Thompson}}, \bibinfo {author}
  {\bibfnamefont {S.~M.}\ \bibnamefont {Girvin}}, \bibinfo {author}
  {\bibfnamefont {A.~A.}\ \bibnamefont {Clerk}}, \bibinfo {author}
  {\bibfnamefont {F.}~\bibnamefont {Marquardt}},\ and\ \bibinfo {author}
  {\bibfnamefont {J.~G.~E.}\ \bibnamefont {Harris}},\ }\href
  {https://doi.org/10.1088/1367-2630/10/9/095008} {\bibfield  {journal}
  {\bibinfo  {journal} {New J. Phys.}\ }\textbf {\bibinfo {volume} {10}},\
  \bibinfo {pages} {095008} (\bibinfo {year} {2008})}\BibitemShut {NoStop}%
\bibitem [{\citenamefont {Zhang}\ \emph {et~al.}(2018)\citenamefont {Zhang},
  \citenamefont {Peng}, \citenamefont {Özdemir}, \citenamefont {Pichler},
  \citenamefont {Krimer}, \citenamefont {Zhao}, \citenamefont {Nori},
  \citenamefont {Liu}, \citenamefont {Rotter},\ and\ \citenamefont
  {Yang}}]{Zhang2018a}%
  \BibitemOpen
  \bibfield  {author} {\bibinfo {author} {\bibfnamefont {J.}~\bibnamefont
  {Zhang}}, \bibinfo {author} {\bibfnamefont {B.}~\bibnamefont {Peng}},
  \bibinfo {author} {\bibfnamefont {a.~K.}\ \bibnamefont {Özdemir}}, \bibinfo
  {author} {\bibfnamefont {K.}~\bibnamefont {Pichler}}, \bibinfo {author}
  {\bibfnamefont {D.~O.}\ \bibnamefont {Krimer}}, \bibinfo {author}
  {\bibfnamefont {G.}~\bibnamefont {Zhao}}, \bibinfo {author} {\bibfnamefont
  {F.}~\bibnamefont {Nori}}, \bibinfo {author} {\bibfnamefont {Y.-x.}\
  \bibnamefont {Liu}}, \bibinfo {author} {\bibfnamefont {S.}~\bibnamefont
  {Rotter}},\ and\ \bibinfo {author} {\bibfnamefont {L.}~\bibnamefont {Yang}},\
  }\href {https://doi.org/10.1038/s41566-018-0213-5} {\bibfield  {journal}
  {\bibinfo  {journal} {Nat. Photonics}\ }\textbf {\bibinfo {volume} {12}},\
  \bibinfo {pages} {479} (\bibinfo {year} {2018})}\BibitemShut {NoStop}%
\bibitem [{\citenamefont {Xu}\ \emph {et~al.}(2024)\citenamefont {Xu},
  \citenamefont {Mao}, \citenamefont {Li}, \citenamefont {Zuo}, \citenamefont
  {Zhang}, \citenamefont {Yang}, \citenamefont {Xu}, \citenamefont {Liu},
  \citenamefont {Deng}, \citenamefont {Chen}, \citenamefont {Xia},
  \citenamefont {Qiu}, \citenamefont {Zhu}, \citenamefont {Jing},\ and\
  \citenamefont {Liu}}]{Xu2024}%
  \BibitemOpen
  \bibfield  {author} {\bibinfo {author} {\bibfnamefont {J.}~\bibnamefont
  {Xu}}, \bibinfo {author} {\bibfnamefont {Y.}~\bibnamefont {Mao}}, \bibinfo
  {author} {\bibfnamefont {Z.}~\bibnamefont {Li}}, \bibinfo {author}
  {\bibfnamefont {Y.}~\bibnamefont {Zuo}}, \bibinfo {author} {\bibfnamefont
  {J.}~\bibnamefont {Zhang}}, \bibinfo {author} {\bibfnamefont
  {B.}~\bibnamefont {Yang}}, \bibinfo {author} {\bibfnamefont {W.}~\bibnamefont
  {Xu}}, \bibinfo {author} {\bibfnamefont {N.}~\bibnamefont {Liu}}, \bibinfo
  {author} {\bibfnamefont {Z.~J.}\ \bibnamefont {Deng}}, \bibinfo {author}
  {\bibfnamefont {W.}~\bibnamefont {Chen}}, \bibinfo {author} {\bibfnamefont
  {K.}~\bibnamefont {Xia}}, \bibinfo {author} {\bibfnamefont {C.-W.}\
  \bibnamefont {Qiu}}, \bibinfo {author} {\bibfnamefont {Z.}~\bibnamefont
  {Zhu}}, \bibinfo {author} {\bibfnamefont {H.}~\bibnamefont {Jing}},\ and\
  \bibinfo {author} {\bibfnamefont {K.}~\bibnamefont {Liu}},\ }\href
  {https://doi.org/10.1038/s41565-024-01729-8} {\bibfield  {journal} {\bibinfo
  {journal} {Nat. Nanotechnol.}\ }\textbf {\bibinfo {volume} {19}},\ \bibinfo
  {pages} {1472} (\bibinfo {year} {2024})}\BibitemShut {NoStop}%
\bibitem [{\citenamefont {Naghiloo}\ \emph {et~al.}(2019)\citenamefont
  {Naghiloo}, \citenamefont {Abbasi}, \citenamefont {Joglekar},\ and\
  \citenamefont {Murch}}]{Naghiloo2019}%
  \BibitemOpen
  \bibfield  {author} {\bibinfo {author} {\bibfnamefont {M.}~\bibnamefont
  {Naghiloo}}, \bibinfo {author} {\bibfnamefont {M.}~\bibnamefont {Abbasi}},
  \bibinfo {author} {\bibfnamefont {Y.~N.}\ \bibnamefont {Joglekar}},\ and\
  \bibinfo {author} {\bibfnamefont {K.~W.}\ \bibnamefont {Murch}},\ }\href
  {https://doi.org/10.1038/s41567-019-0652-z} {\bibfield  {journal} {\bibinfo
  {journal} {Nat. Phys.}\ }\textbf {\bibinfo {volume} {15}},\ \bibinfo {pages}
  {1232} (\bibinfo {year} {2019})}\BibitemShut {NoStop}%
\bibitem [{\citenamefont {Liu}\ \emph {et~al.}(2021)\citenamefont {Liu},
  \citenamefont {Wu}, \citenamefont {Duan}, \citenamefont {Rong},\ and\
  \citenamefont {Du}}]{Liu2021}%
  \BibitemOpen
  \bibfield  {author} {\bibinfo {author} {\bibfnamefont {W.}~\bibnamefont
  {Liu}}, \bibinfo {author} {\bibfnamefont {Y.}~\bibnamefont {Wu}}, \bibinfo
  {author} {\bibfnamefont {C.-K.}\ \bibnamefont {Duan}}, \bibinfo {author}
  {\bibfnamefont {X.}~\bibnamefont {Rong}},\ and\ \bibinfo {author}
  {\bibfnamefont {J.}~\bibnamefont {Du}},\ }\href
  {https://doi.org/10.1103/PhysRevLett.126.170506} {\bibfield  {journal}
  {\bibinfo  {journal} {Phys. Rev. Lett.}\ }\textbf {\bibinfo {volume} {126}},\
  \bibinfo {pages} {170506} (\bibinfo {year} {2021})}\BibitemShut {NoStop}%
\bibitem [{\citenamefont {Chen}\ \emph {et~al.}(2021)\citenamefont {Chen},
  \citenamefont {Abbasi}, \citenamefont {Joglekar},\ and\ \citenamefont
  {Murch}}]{Chen2021}%
  \BibitemOpen
  \bibfield  {author} {\bibinfo {author} {\bibfnamefont {W.}~\bibnamefont
  {Chen}}, \bibinfo {author} {\bibfnamefont {M.}~\bibnamefont {Abbasi}},
  \bibinfo {author} {\bibfnamefont {Y.~N.}\ \bibnamefont {Joglekar}},\ and\
  \bibinfo {author} {\bibfnamefont {K.~W.}\ \bibnamefont {Murch}},\ }\href
  {https://doi.org/10.1103/PhysRevLett.127.140504} {\bibfield  {journal}
  {\bibinfo  {journal} {Phys. Rev. Lett.}\ }\textbf {\bibinfo {volume} {127}},\
  \bibinfo {pages} {140504} (\bibinfo {year} {2021})}\BibitemShut {NoStop}%
\bibitem [{\citenamefont {Lu}\ \emph {et~al.}(2025)\citenamefont {Lu},
  \citenamefont {Li},\ and\ \citenamefont {Wang}}]{Lu2025}%
  \BibitemOpen
  \bibfield  {author} {\bibinfo {author} {\bibfnamefont {Y.-W.}\ \bibnamefont
  {Lu}}, \bibinfo {author} {\bibfnamefont {W.}~\bibnamefont {Li}},\ and\
  \bibinfo {author} {\bibfnamefont {X.-H.}\ \bibnamefont {Wang}},\ }\href
  {https://doi.org/10.1021/acsnano.4c15648} {\bibfield  {journal} {\bibinfo
  {journal} {ACS Nano}\ }\textbf {\bibinfo {volume} {19}},\ \bibinfo {pages}
  {17953} (\bibinfo {year} {2025})}\BibitemShut {NoStop}%
\bibitem [{\citenamefont {Lee}\ \emph {et~al.}(2019)\citenamefont {Lee},
  \citenamefont {Lee}, \citenamefont {Hong}, \citenamefont {Sch\"uppert},
  \citenamefont {Kwon}, \citenamefont {Kim}, \citenamefont {Colombe},
  \citenamefont {Northup}, \citenamefont {Cho},\ and\ \citenamefont
  {Blatt}}]{Lee2019a}%
  \BibitemOpen
  \bibfield  {author} {\bibinfo {author} {\bibfnamefont {M.}~\bibnamefont
  {Lee}}, \bibinfo {author} {\bibfnamefont {M.}~\bibnamefont {Lee}}, \bibinfo
  {author} {\bibfnamefont {S.}~\bibnamefont {Hong}}, \bibinfo {author}
  {\bibfnamefont {K.}~\bibnamefont {Sch\"uppert}}, \bibinfo {author}
  {\bibfnamefont {Y.-D.}\ \bibnamefont {Kwon}}, \bibinfo {author}
  {\bibfnamefont {T.}~\bibnamefont {Kim}}, \bibinfo {author} {\bibfnamefont
  {Y.}~\bibnamefont {Colombe}}, \bibinfo {author} {\bibfnamefont {T.~E.}\
  \bibnamefont {Northup}}, \bibinfo {author} {\bibfnamefont {D.-I.~D.}\
  \bibnamefont {Cho}},\ and\ \bibinfo {author} {\bibfnamefont {R.}~\bibnamefont
  {Blatt}},\ }\href {https://doi.org/10.1103/PhysRevApplied.12.044052}
  {\bibfield  {journal} {\bibinfo  {journal} {Phys. Rev. Applied}\ }\textbf
  {\bibinfo {volume} {12}},\ \bibinfo {pages} {044052} (\bibinfo {year}
  {2019})}\BibitemShut {NoStop}%
\bibitem [{\citenamefont {Ong}\ \emph {et~al.}(2020)\citenamefont {Ong},
  \citenamefont {Sch\"uppert}, \citenamefont {Jobez}, \citenamefont {Teller},
  \citenamefont {Ames}, \citenamefont {Fioretto}, \citenamefont {Friebe},
  \citenamefont {Lee}, \citenamefont {Colombe}, \citenamefont {Blatt},\ and\
  \citenamefont {Northup}}]{Ong2020}%
  \BibitemOpen
  \bibfield  {author} {\bibinfo {author} {\bibfnamefont {F.~R.}\ \bibnamefont
  {Ong}}, \bibinfo {author} {\bibfnamefont {K.}~\bibnamefont {Sch\"uppert}},
  \bibinfo {author} {\bibfnamefont {P.}~\bibnamefont {Jobez}}, \bibinfo
  {author} {\bibfnamefont {M.}~\bibnamefont {Teller}}, \bibinfo {author}
  {\bibfnamefont {B.}~\bibnamefont {Ames}}, \bibinfo {author} {\bibfnamefont
  {D.~A.}\ \bibnamefont {Fioretto}}, \bibinfo {author} {\bibfnamefont
  {K.}~\bibnamefont {Friebe}}, \bibinfo {author} {\bibfnamefont
  {M.}~\bibnamefont {Lee}}, \bibinfo {author} {\bibfnamefont {Y.}~\bibnamefont
  {Colombe}}, \bibinfo {author} {\bibfnamefont {R.}~\bibnamefont {Blatt}},\
  and\ \bibinfo {author} {\bibfnamefont {T.~E.}\ \bibnamefont {Northup}},\
  }\href@noop {} {\bibfield  {journal} {\bibinfo  {journal} {New J. Phys.}\
  }\textbf {\bibinfo {volume} {22}},\ \bibinfo {pages} {063018} (\bibinfo
  {year} {2020})}\BibitemShut {NoStop}%
\bibitem [{\citenamefont {Takahashi}\ \emph {et~al.}(2020)\citenamefont
  {Takahashi}, \citenamefont {Kassa}, \citenamefont {Christoforou},\ and\
  \citenamefont {Keller}}]{Takahashi2020}%
  \BibitemOpen
  \bibfield  {author} {\bibinfo {author} {\bibfnamefont {H.}~\bibnamefont
  {Takahashi}}, \bibinfo {author} {\bibfnamefont {E.}~\bibnamefont {Kassa}},
  \bibinfo {author} {\bibfnamefont {C.}~\bibnamefont {Christoforou}},\ and\
  \bibinfo {author} {\bibfnamefont {M.}~\bibnamefont {Keller}},\ }\href
  {https://doi.org/10.1103/PhysRevLett.124.013602} {\bibfield  {journal}
  {\bibinfo  {journal} {Phys. Rev. Lett.}\ }\textbf {\bibinfo {volume} {124}},\
  \bibinfo {pages} {013602} (\bibinfo {year} {2020})}\BibitemShut {NoStop}%
\bibitem [{\citenamefont {Sounas}(2020)}]{Sounas2020}%
  \BibitemOpen
  \bibfield  {author} {\bibinfo {author} {\bibfnamefont {D.~L.}\ \bibnamefont
  {Sounas}},\ }\href {https://doi.org/10.1103/PhysRevB.101.104303} {\bibfield
  {journal} {\bibinfo  {journal} {Phys. Rev. B}\ }\textbf {\bibinfo {volume}
  {101}},\ \bibinfo {pages} {104303} (\bibinfo {year} {2020})}\BibitemShut
  {NoStop}%
\end{thebibliography}%


\begin{thebibliography}{3}%
\makeatletter
\providecommand \@ifxundefined [1]{%
 \@ifx{#1\undefined}
}%
\providecommand \@ifnum [1]{%
 \ifnum #1\expandafter \@firstoftwo
 \else \expandafter \@secondoftwo
 \fi
}%
\providecommand \@ifx [1]{%
 \ifx #1\expandafter \@firstoftwo
 \else \expandafter \@secondoftwo
 \fi
}%
\providecommand \natexlab [1]{#1}%
\providecommand \enquote  [1]{``#1''}%
\providecommand \bibnamefont  [1]{#1}%
\providecommand \bibfnamefont [1]{#1}%
\providecommand \citenamefont [1]{#1}%
\providecommand \href@noop [0]{\@secondoftwo}%
\providecommand \href [0]{\begingroup \@sanitize@url \@href}%
\providecommand \@href[1]{\@@startlink{#1}\@@href}%
\providecommand \@@href[1]{\endgroup#1\@@endlink}%
\providecommand \@sanitize@url [0]{\catcode `\\12\catcode `\$12\catcode
  `\&12\catcode `\#12\catcode `\^12\catcode `\_12\catcode `\%12\relax}%
\providecommand \@@startlink[1]{}%
\providecommand \@@endlink[0]{}%
\providecommand \url  [0]{\begingroup\@sanitize@url \@url }%
\providecommand \@url [1]{\endgroup\@href {#1}{\urlprefix }}%
\providecommand \urlprefix  [0]{URL }%
\providecommand \Eprint [0]{\href }%
\providecommand \doibase [0]{https://doi.org/}%
\providecommand \selectlanguage [0]{\@gobble}%
\providecommand \bibinfo  [0]{\@secondoftwo}%
\providecommand \bibfield  [0]{\@secondoftwo}%
\providecommand \translation [1]{[#1]}%
\providecommand \BibitemOpen [0]{}%
\providecommand \bibitemStop [0]{}%
\providecommand \bibitemNoStop [0]{.\EOS\space}%
\providecommand \EOS [0]{\spacefactor3000\relax}%
\providecommand \BibitemShut  [1]{\csname bibitem#1\endcsname}%
\let\auto@bib@innerbib\@empty
\bibitem [{\citenamefont {Johnson}\ and\ \citenamefont
  {Christy}(1972)}]{Johnson1972}%
  \BibitemOpen
  \bibfield  {author} {\bibinfo {author} {\bibfnamefont {P.~B.}\ \bibnamefont
  {Johnson}}\ and\ \bibinfo {author} {\bibfnamefont {R.~W.}\ \bibnamefont
  {Christy}},\ }\bibfield  {title} {\bibinfo {title} {Optical constants of the
  noble metals},\ }\href {https://doi.org/10.1103/PhysRevB.6.4370} {\bibfield
  {journal} {\bibinfo  {journal} {Phys. Rev. B}\ }\textbf {\bibinfo {volume}
  {6}},\ \bibinfo {pages} {4370} (\bibinfo {year} {1972})}\BibitemShut
  {NoStop}%
\bibitem [{\citenamefont {Agrawal}(2013)}]{Agrawal2013b}%
  \BibitemOpen
  \bibfield  {author} {\bibinfo {author} {\bibfnamefont {G.}~\bibnamefont
  {Agrawal}},\ }\href {https://books.google.co.kr/books?id=xNvw-GDVn84C} {\emph
  {\bibinfo {title} {Nonlinear Fiber Optics}}},\ Optics and Photonics\
  (\bibinfo  {publisher} {Elsevier Science},\ \bibinfo {year}
  {2013})\BibitemShut {NoStop}%
\bibitem [{\citenamefont {Ding}\ \emph {et~al.}(2022)\citenamefont {Ding},
  \citenamefont {Fang},\ and\ \citenamefont {Ma}}]{Ding2022}%
  \BibitemOpen
  \bibfield  {author} {\bibinfo {author} {\bibfnamefont {K.}~\bibnamefont
  {Ding}}, \bibinfo {author} {\bibfnamefont {C.}~\bibnamefont {Fang}},\ and\
  \bibinfo {author} {\bibfnamefont {G.}~\bibnamefont {Ma}},\ }\bibfield
  {title} {\bibinfo {title} {Non-hermitian topology and exceptional-point
  geometries},\ }\href {https://doi.org/10.1038/s42254-022-00516-5} {\bibfield
  {journal} {\bibinfo  {journal} {Nat. Rev. Phys.}\ }\textbf {\bibinfo {volume}
  {4}},\ \bibinfo {pages} {745} (\bibinfo {year} {2022})}\BibitemShut {NoStop}%
\end{thebibliography}%

\end{document}


\title{Supplemental Material: In-situ tuning of cavity dissipation and a topological transition in an atom-nanotip-cavity system} 
\author{Taegyu Ha}
\thanks{These authors contributed equally to this work.}
\affiliation{Department of Electrical Engineering, Pohang University of Science and Technology (POSTECH), 37673 Pohang, Korea}
\author{Kiyanoush Goudarzi$^{\dagger ,}$} 
\thanks{These authors contributed equally to this work.}
\affiliation{Department of Electrical Engineering, Pohang University of Science and Technology (POSTECH), 37673 Pohang, Korea}
\author{Dowon Lee}
\affiliation{Department of Electrical Engineering, Pohang University of Science and Technology (POSTECH), 37673 Pohang, Korea}
\author{Donggeon Kim}
\affiliation{Department of Electrical Engineering, Pohang University of Science and Technology (POSTECH), 37673 Pohang, Korea}
\author{Eunchul Jeong}
\affiliation{Graduate School of Convergence Science and Technology, Pohang University of Science and Technology (POSTECH), 37673 Pohang, Korea}
\author{Uijin Kim}
\affiliation{Department of Electrical Engineering, Pohang University of Science and Technology (POSTECH), 37673 Pohang, Korea}
\author{Myunghun Kim}
\affiliation{Department of Electrical Engineering, Pohang University of Science and Technology (POSTECH), 37673 Pohang, Korea}
\author{Jinuk Kim}
\affiliation{Quantum Technology Institute, Korea Research Institute of Standards and Science (KRISS), 34113 Daejeon, Korea}
\author{Moonjoo Lee}
\email{kgoudarzi@postech.ac.kr $and$ moonjoo.lee@postech.ac.kr}
\affiliation{Department of Electrical Engineering, Pohang University of Science and Technology (POSTECH), 37673 Pohang, Korea}

\date{\today}

\maketitle

\section{Numerical calculation of electric fields in the nanotip-cavity system}

All numerical simulations of the electric field are performed using COMSOL Multiphysics, employing the Wave Optics Module in the frequency domain.
We consider a configuration where a gold nanotip is inserted into the cavity mode. 
The cavity is formed by left and right multilayer dielectric mirrors, with their refractive indices and thicknesses described in the main text.
Other material's property, the refractive index of gold, $n_{\textrm{g}}$, is taken from Ref.~\cite{Johnson1972}.
We exploit the interpolated values of Re$(n_{\textrm{g}}) = 0.1474$ and Im$(n_{\textrm{g}})= 4.7414$ at $780$~nm.
\begin{figure}[!b]
	\begin{center}
		\includegraphics[width=0.5 \linewidth]{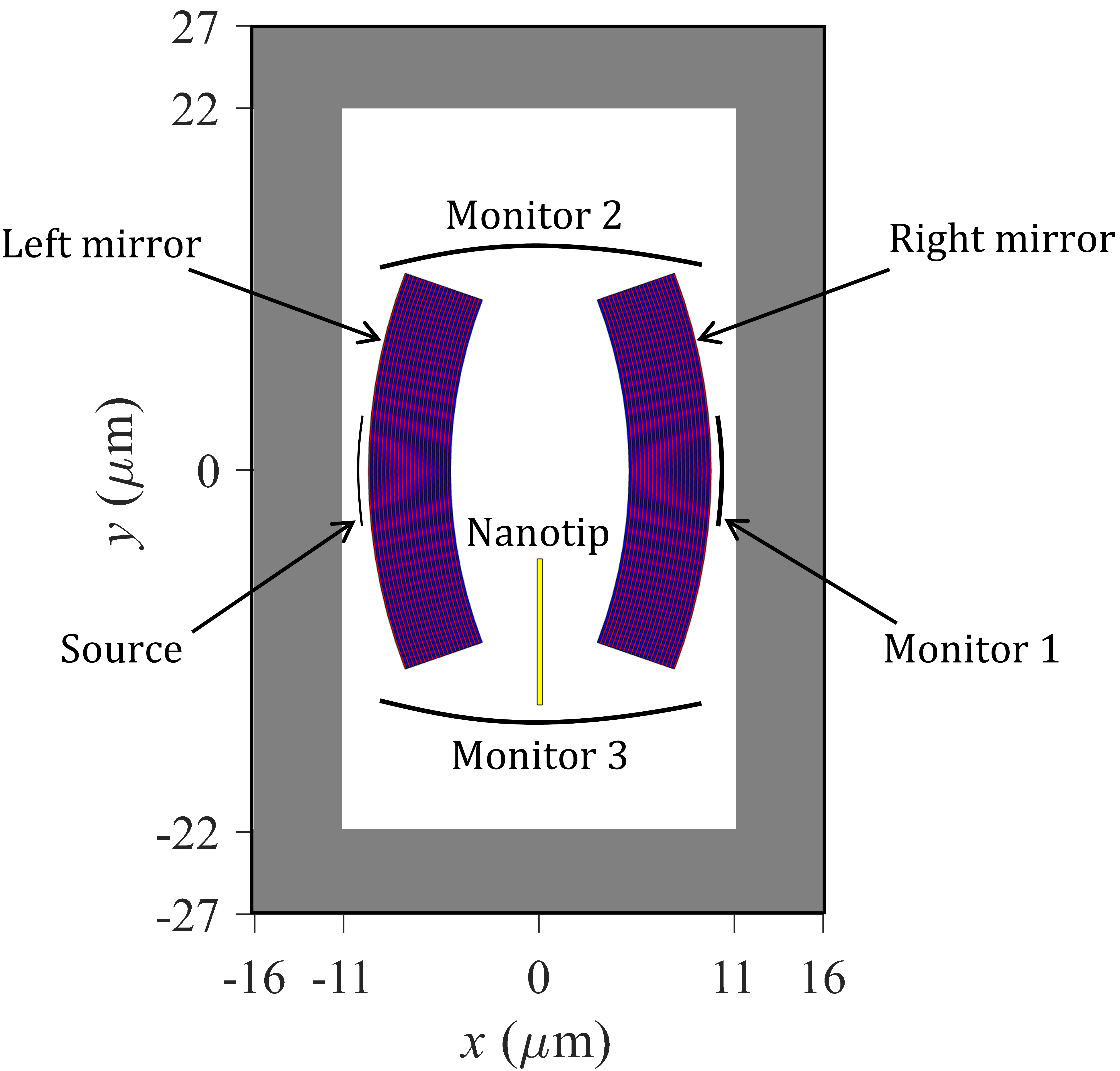}
		\caption{
			Schematic of the cavity and nanotip used for two-dimensional numerical simulations.
			Multilayer coating is depicted in blue and red; gray region represents the perfectly matched layer.		
		}
	\label{fig:schematic}
	\end{center}
\end{figure}

\begin{figure*}[!t]
	\begin{center}
		\includegraphics[width=1 \linewidth]{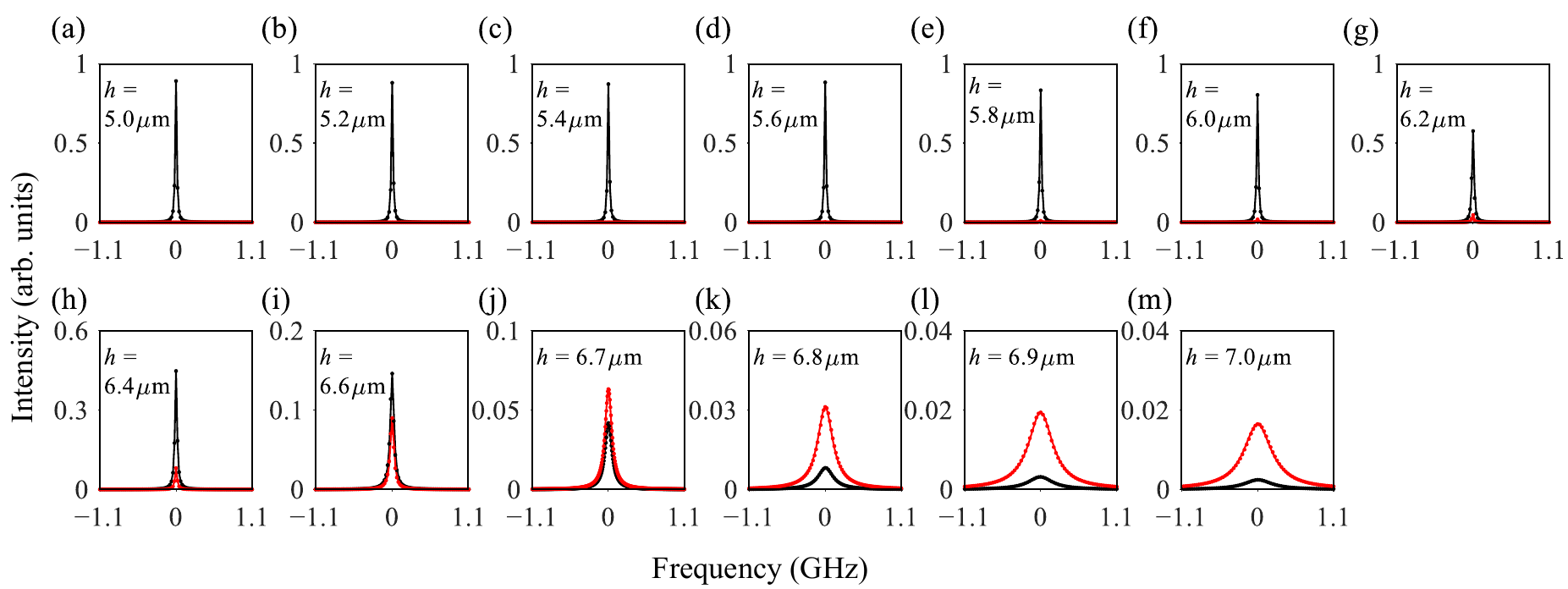}
		\caption{
			(a)--(m) Calculation results of the intensity at Monitor 1 and 2, for several tip positions ranging from $h=5.0$ to $7.0$~$\mu$m.
			Each `$0$' on the horizontal axis denotes the center frequency of the corresponding peak. 
			Each simulation run takes approximately $15$~minutes. 
			Calculation results are shown in points, and the fitting results with Lorentzian functions are in black (Monitor $1$) and red (Monitor $2$).
			Detected signal at Monitor 3 is on the order of $10^{-4}$ maximally.
			}	
			\label{fig:spectrum}
	\end{center}
\end{figure*}

In our simulation, the driving field is introduced at a source, exciting the TEM$_{00}$ mode of the cavity, and the transmitted signal is subsequently recorded at Monitor 1 (Fig.~\ref{fig:schematic}), and scattered signal is at Monitor 2 and 3.
The accuracy and precision of our calculation are primarily limited by the finite size of the simulation area: Backreflections at the domain boundaries interfere with the cavity field, leading to distortions---most notably in the calculated cavity resonance frequency. 
To suppress these reflections, a perfectly matched layer (PML), shown in gray in Fig.~\ref{fig:schematic}, is used to surround the simulation area and mitigate boundary effects. 
In the PML, the electric field decays exponentially with a decay constant of $770 \sim 870$~nm, and the reflectivity is on the order of $10^{-8}$.
However, this field still interferes with the cavity field, governing the errors of our calculation.  
Therefore, we conduct $6$ calculations for different sizes of PML, take the average and standard deviation of the results, and plot the results in Figs.~2(g)--(i).
The obtained intensity spectrum is fit to the Lorentzian function
\begin{equation}
		L(\omega) = \frac{A\kappa^{2}}{\kappa^{2} + \left( \omega - \omega_{c} \right)^{2}},
	\label{eq:Lorentzian}
\end{equation}

\noindent
where $A$ is the amplitude, $\omega_{\textrm{c}}$ is the center frequency of the spectrum, and $\kappa$ is the cavity decay rate. 
The calculation and fitting results are shown in Fig.~\ref{fig:spectrum} and also in Figs.~2(d)--(f).  
\begin{figure}[!t]
	\begin{center}
		\includegraphics[width=1 \linewidth]{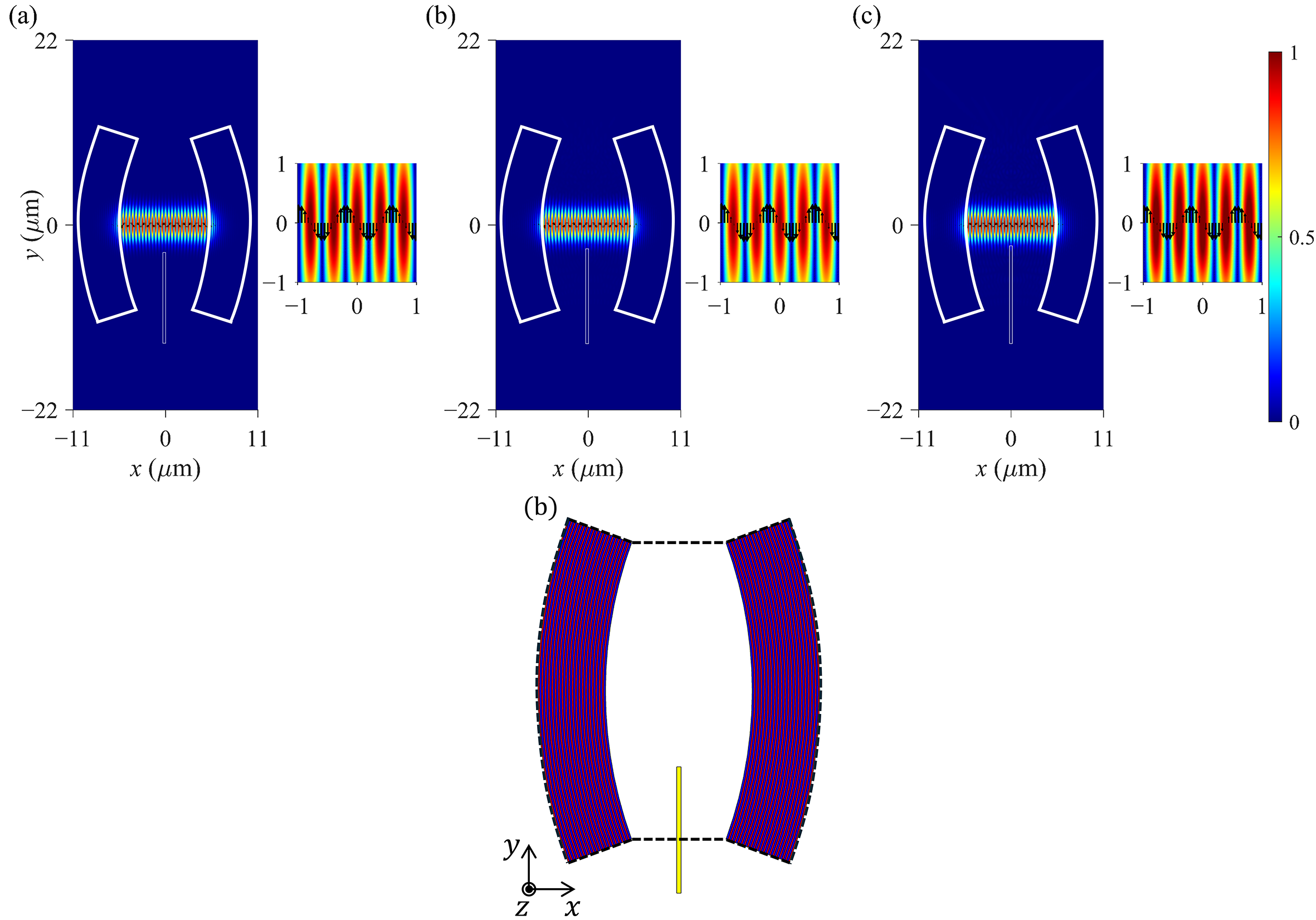} 
		\caption{
			(a)--(c) Normalized electric field distribution for $h=5$, $6.8$, and $7$~$\mu$m, respectively. 
			The color scale represents the absolute value of the electric field amplitude.
			(inset) Magnified view of selected region.
			Arrows indicate the direction of the magnetic field, with their length corresponding to the field amplitude.
			(d) The black dashed line outlines the region over which the cavity field is integrated to calculate $A_{\textrm{eff}}$.	
			}
		\label{fig:field_distribution}
	\end{center}
\end{figure}


\section{Calculation of atom-cavity coupling constant}

We define the maximum atom-cavity coupling constant
\begin{equation}		
		g_{0} = \sqrt{\frac{\omega_{0}} {2 \hbar \left( \sum_{i=0}^2 \varepsilon_{i} A_{\textrm{eff}, i}   \right)  \sqrt{\pi} w_{0} }} d_{\textrm{ge}},
	\label{eq:g_0}
\end{equation}

\noindent 
where $\omega_{0}$ is the atomic transition frequency, $\hbar$ is the Planck constant, $\epsilon_{0}$ is the vacuum permittivity, $A_{\textrm{eff}, 0}$ is the mode area in vacuum, $\epsilon_{1}$ is the permittivity of SiO$_{2}$, $A_{\textrm{eff}, 1}$ is the mode area in SiO$_{2}$, $\epsilon_{2}$ is the permittivity of Ta$_{2}$O$_{5}$, $A_{\textrm{eff}, 2}$ is the mode area in Ta$_{2}$O$_{5}$, $w_{0}$ is the mode waist of $1.70(6)$~$\mu$m, $d_{\textrm{ge}} = 3.584\times 10^{-29} / \sqrt{2}$ C$\cdot$m is the dipole moment of the atomic transition of $^{87}$Rb.
We calculate $A_{\textrm{eff}, i}$ as follows~\cite{Agrawal2013b}
\begin{equation}
	\begin{aligned}
		A_{\text{eff}, i} = \frac{\left( \int \int |\textrm{E}_{i}(x, y)|^2 \, dx \, dy \right)^2}{\int \int |\textrm{E}_{i}(x, y)|^4 \, dx \, dy},		
	\end{aligned}
	\label{eq:mode_area}
\end{equation}

\noindent 
where $\textrm{E}_{0}(x, y)$ is the electric field distribution in vacuum, $\textrm{E}_{1}(x, y)$ is in SiO$_{2}$, and $\textrm{E}_{2}(x, y)$ is in Ta$_{2}$O$_{5}$, respectively. 
We obtain $A_{\textrm{eff,0}} = 2.09(2) \times 10^{-11}$~m$^{2}$, $A_{\textrm{eff,1}} = 2.00(3) \times 10^{-12}$~m$^{2}$, and $A_{\textrm{eff,2}} = 1.24(5) \times 10^{-12}$~m$^{2}$, which hardly change from $h=5$ to $7$~$\mu$m.
Consequently, Eq.~\eqref{eq:g_0} yields $g_0 / (2\pi) = 480$~MHz.

\section{Hamiltonian and quantum Monte Carlo simulation}

The Hamiltonian of the driven atom-cavity system is given by
\begin{equation}
		H = \omega_{\textrm{a}} \sigma_{+}\sigma_{-} + \omega_{\textrm{c}} a^{\dagger}a +  g (a^{\dagger} \sigma_{-} + \sigma_{+}a) + \epsilon(a e^{-i\omega_{\textrm{p}} t} + a^{\dagger}e^{i\omega_{\textrm{p}} t}),
	\label{eq:Hamiltonian}
\end{equation}
\noindent 
where $\sigma_{+} = | e \rangle \langle g |$ and $\sigma_{-} = | g \rangle \langle e |$ represent the atomic raising and lowering operators, respectively. 
The operators $a$ and $a^{\dagger}$ correspond to the annihilation and creation operators of the cavity field. 
The parameters $\omega_{\textrm{a}}$, $\omega_{\textrm{c}}$, and $\omega_{\textrm{p}}$ denote the atomic, cavity, and pump frequencies, respectively, while $g$ represents the atom-cavity coupling constant, $\epsilon$ is the amplitude of the probe field, and $\hbar = 1$.

In order to obtain the Hamiltonian in the interaction picture, we define the free Hamiltonian as
\begin{equation}
	H_{0} = \omega_{\textrm{p}} (\sigma_{+} \sigma_{-} + a^{\dagger}a),
\end{equation}

\noindent
such that the total Hamiltonian of Eq.~\eqref{eq:Hamiltonian} can be expressed as $H = H_0 + H_1$. 
The Hamiltonian in the interaction picture is then given by
\begin{align}
	H_{\text{int}} & = e^{iH_{0} t} H_1 e^{-iH_{0} t} \\
		                  & = (\omega_a - \omega_p) \sigma_{+}\sigma_{-} + (\omega_c - \omega_p) a^{\dagger}a+ g( a^{\dagger}\sigma_{-} + \sigma_{+}a) + \epsilon (a + a^{\dagger}). \label{eq:H_int}	
\end{align}

\noindent
Utilizing the QuTip package in Python, we numercally solve the master equation with the interaction Hamiltonian given in Eq.~\eqref{eq:H_int} to perform quantum Monte Carlo (QMC) simulations presented in the main text.
The parameters used for the QMC simulations are as follows:
\begin{itemize}
	\item Spontaneous emission rate: $\gamma = 2\pi \times 3.03$ MHz,
	\item Probe field amplitude: $\epsilon = 2\pi \times 1.0$ MHz,
	\item Maximum Fock state truncation: $N = 3$,
	\item Mean photon number range: $[0.00097, 0.0087]$, ensuring the weak excitation regime.
\end{itemize}

\section{Extracting the eigenvalues of the non-Hermitian matrix}

To extract the eigenvalues of the non-Hermitian Hamiltonian, we derive an analytical expression for the cavity transmission spectrum.

\subsection{Non-Hermitian Hamiltonian}
 
We begin with the non-Hermitian Hamiltonian for the atom-cavity system
\begin{equation}
	H_{\mathrm{nH}} = 
	\begin{pmatrix}
		\omega_{\textrm{a}} - i\gamma & g \\
		g & \omega_{\textrm{c}} - i\kappa
	\end{pmatrix},
	\label{eq:HnH}
\end{equation}

\noindent
where $\omega_a$ and $\omega_c$ denote the atomic and cavity resonance frequencies, respectively. The parameters $\gamma$ and $\kappa$ represent the decay rates of the atom and the cavity, while $g$ is the coupling strength. Defining the detuning $\Delta = \omega_{\textrm{c}} - \omega_{\textrm{a}}$, we rewrite the non-Hermitian Hamiltonian in a shifted frame as
\begin{equation}
	H'_{\mathrm{nH}} = 
	\begin{pmatrix}
		A & g \\
		g & B
	\end{pmatrix},
	\label{eq:HnH_shifted}
\end{equation}

\noindent
with $A = -i\gamma$ and $B = \Delta - i\kappa$.

\subsection{Decomposition matrix for the non-Hermitian Hamiltonian}

The eigenvalues of the reduced Hamiltonian $H'_{\mathrm{nH}}$ are given by
\begin{equation}
	\lambda_{\pm} = \frac{A + B \pm \sqrt{(A - B)^2 + 4g^2}}{2},
	\label{eq:eigenvalues}
\end{equation}
\noindent
with the eigenvectors of 
\begin{equation}
	\begin{pmatrix}
		c_{1,\pm} \\
		c_{2,\pm}
	\end{pmatrix}
	=
	c_{\textrm{n}}
	\begin{pmatrix}
		g \\
		\lambda_{\pm} - A
	\end{pmatrix}.
	\label{eq:eigenvectors}
\end{equation}
where the normalization constant $c_{\textrm{n}}$.
We then obtain the decomposition matrix of eigenvectors as
\begin{equation}
	\Theta = 
	\begin{pmatrix}
		g & g \\
		\lambda_{+} - A & \lambda_{-} - A
	\end{pmatrix},
	\label{eq:Theta}
\end{equation}
and the diagonal matrix of eigenvalues
\begin{equation}
	\Lambda = 
	\begin{pmatrix}
		\lambda_{+} & 0 \\
		0 & \lambda_{-}
	\end{pmatrix}.
	\label{eq:Lambda}
\end{equation}
This yields another representation of the Hamiltonian:
\begin{equation}
	H'_{\mathrm{nH}} = \Theta \Lambda \Theta^{-1}.
	\label{eq:reconstructed_H}
\end{equation}

\begin{figure}[!t]
	\begin{center}
		\includegraphics[width=1 \linewidth]{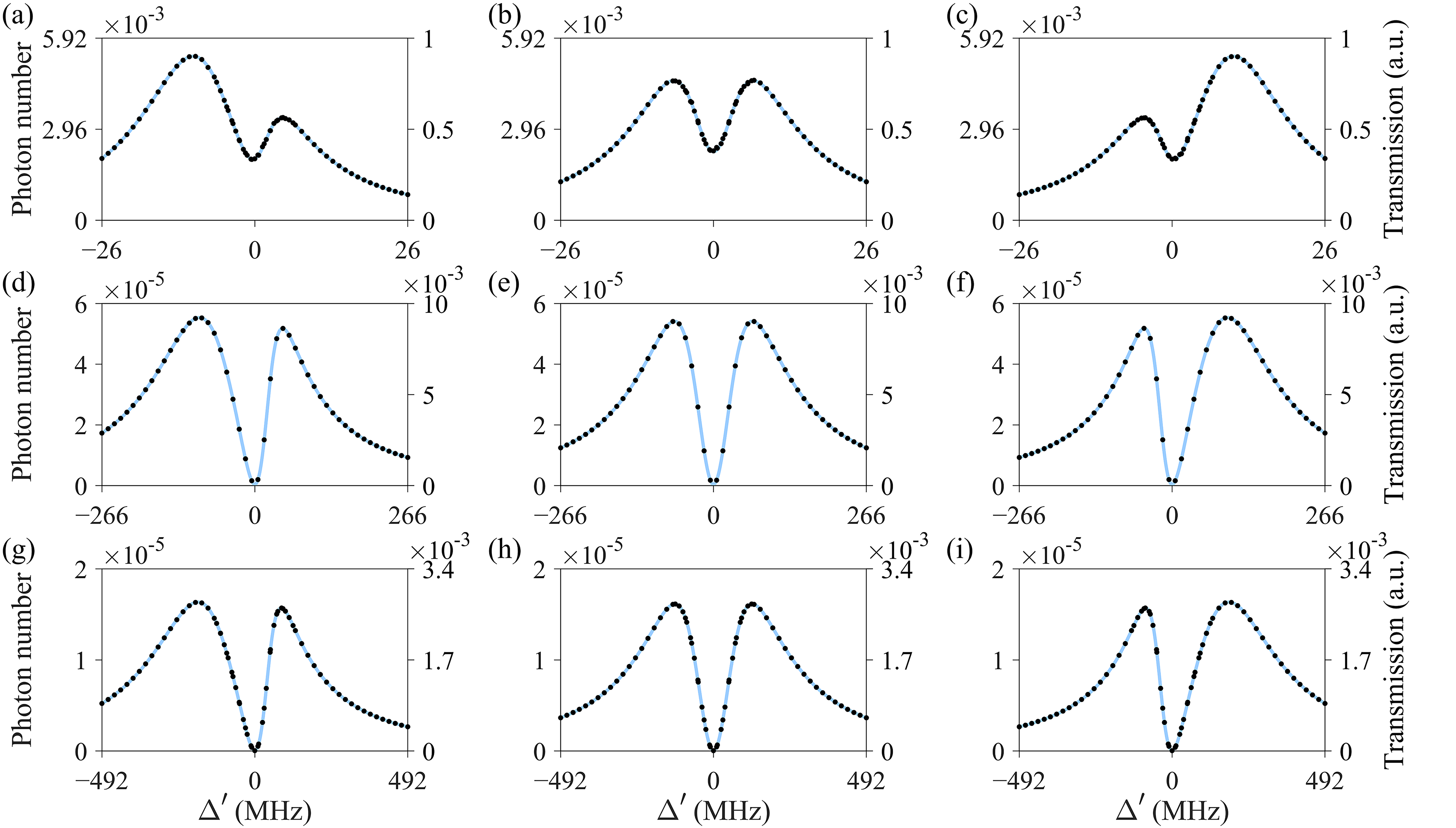} 
		\caption{
			Cavity transmission as a function of $\Delta^{\prime}$ for comparison between QMC simulation results and fitting with Eq.~\eqref{eq:transmission}.
			Black points are the results of QMC simulations, and the light blue lines are the fitting results. 
			Top, center, and bottom rows correspond to ($\kappa$, $g$)$/(2\pi)$ of $(13, 5)$, $(133, 65)$, and $(246, 121.5)$~MHz, respectively. 
			$\Delta_{\textrm{ca}}=(-7, 0, 7)$~MHz for (a)--(c), $\Delta_{\textrm{ca}}= -50, 0, 50$~MHz for (d)--(f), $\Delta_{\textrm{ca}}= -100, 0,
			100$~MHz for (g)--(i), respectively.}
		\label{fig:EP fitting}
	\end{center}
\end{figure}

\subsection{Analytic expression of cavity transmission}

We now consider the time evolution of the system driven by a weak probe field of an amplitude $\epsilon \ll \kappa$. 
In the single-excitation subspace, the wavefunction is given by
\begin{equation}
	|\psi(t)\rangle = \alpha(t)|e,0\rangle + \beta(t)|g,1\rangle,
\end{equation}
\noindent
which evolves according to the equation of motion
\begin{equation}
	i\frac{\partial\psi(t)\rangle}{\partial t} = (H'_{\mathrm{nH}} - \Delta' I)|\psi(t)\rangle + 
	\begin{pmatrix}
		0 \\
		\epsilon
	\end{pmatrix},
	\label{eq:eom}
\end{equation}
where $\Delta' = \omega_{\textrm{p}} - \omega_{\textrm{c}}$ is the probe detuning and $I$ is the identity matrix.
In the steady state, the system satisfies
\begin{equation}
	\begin{pmatrix}
		\alpha_{\textrm{ss}} \\
		\beta_{\textrm{ss}}
	\end{pmatrix}
	= (H'_{\mathrm{nH}} - \Delta' I)^{-1}
	\begin{pmatrix}
		0 \\
		- \epsilon
	\end{pmatrix}.
	\label{eq:steady_state}
\end{equation}
Using the eigen-decomposition of $H'_{\mathrm{nH}}$, the inverse operator can be expressed as
\begin{equation}
	(H'_{\mathrm{nH}} - \Delta' I)^{-1} = \Theta (\Lambda - \Delta' I)^{-1} \Theta^{-1}.
	\label{eq:inverse_decomposition}
\end{equation}
Substituting into Eq.~\eqref{eq:steady_state}, we find an explicit expression for $\beta_{ss}$:
\begin{equation}
	\beta_{ss} = \epsilon \frac{\Delta' - A}{(\lambda_{-} - \Delta')(\lambda_{+} - \Delta')}.
	\label{eq:beta_ss}
\end{equation}
We thus obtain the cavity transmission spectrum:
\begin{equation}
	T = \left| \frac{\beta_{ss}}{\varepsilon/\kappa} \right|^2 
	= \left| \frac{\kappa(\Delta' - A)}{(\lambda_{-} - \Delta')(\lambda_{+} - \Delta')} \right|^2.
	\label{eq:transmission}
\end{equation}

\noindent
The fitting results for several $g$ and $\kappa$ are shown in Fig.~\ref{fig:EP fitting}.

\section{Scaling behavior near the exceptional line}

We rewrite the non-Hermitian Hamiltonian 
\begin{equation}
	\begin{aligned}
		H_{\textrm{nH}} = \begin{pmatrix}
			F & g \\
			g & G
		\end{pmatrix},
	\end{aligned}
	\label{eq:Eq.8}
\end{equation}
\noindent 
where $F = \omega_{\textrm{a}} - i\gamma$ and $G = \omega_{\textrm{c}} - i\kappa$. 
The characteristic equation of the Hamiltonian is
\begin{equation}
	E^{2} - (F + G)E + FG - g^{2} = 0, 
\end{equation}
resulting in the eigenvalues
\begin{equation}
	E_{\pm} = \frac{(F + G) \pm \sqrt{\delta}}{2},
\end{equation}

\noindent
with $\delta = (F - G)^{2} + 4 g^{2}$.
The exceptional line (EL) occurs when $\delta = 0$, leading to
\begin{equation}
	E_{\mathrm{EL}} = \frac{F + G}{2}.
\end{equation}

\noindent
By applying a small perturbation to $g$ \text{ and } $\kappa$, we can show the behavior of the eigenvalues around the exceptional line.
We first consider a perturbation to $\kappa$ around the EL, near $g = g_{\mathrm{EL}}$ and $\kappa = \kappa_{\mathrm{EL}}$, and $\epsilon \ll 1$. 
Using a perturbation theory, the eigenvalues take the form of
\begin{equation}
	\begin{aligned}
		E = E_{\mathrm{EL}} + c_{1}\epsilon^{1/2} + c_{2}\epsilon.
	\end{aligned}
	\label{eq:perturbation}
\end{equation}

\noindent
The characteristic equation under the perturbation theory is then given by
\begin{equation}
		E^2 \added{-} E(F + G_{\mathrm{EL}} + \epsilon) + F(G_{\mathrm{EL}} + \epsilon) - g_{\mathrm{EL}}^2 = 0,
\end{equation}

\noindent
which is further reduced to 
\begin{equation}
		(E - E_{\mathrm{EL}})^2 \added{-} E\epsilon + \epsilon F = 0,
	\label{eq:reduced_ch_eq}
\end{equation}

\noindent 
where $G_{\mathrm{EL}} = \omega_{\textrm{c}} - i\kappa_{\mathrm{EL}}$.
Substituting Eq.~\eqref{eq:perturbation} into Eq.~\eqref{eq:reduced_ch_eq} results in
\begin{equation}
	c_1^2\epsilon + c_2^2\epsilon^2 + 2 c_1 c_2\epsilon^{3/2} \added{-} E_{\mathrm{EL}}\epsilon \added{-} c_1\epsilon^{3/2} \added{-} c_2\epsilon^{2} + \epsilon F = 0.
	\label{eq:Eq.14}
\end{equation}

\noindent
By equating the coefficients of $\epsilon$ and $\epsilon^{3/2}$, the eigenvalues become
\begin{equation}
	E_{\pm} = \pm \sqrt{\frac{(F_{\mathrm{EL}} - G)}{2}}\epsilon^{1/2} \added{+} \frac{1}{2}\epsilon.
	\label{eq:Eq.15}
\end{equation}

\noindent
Consequently, the energy difference is
\begin{equation}
	\Delta E = \sqrt{2(F_{\mathrm{EL}} - G)}\epsilon^{1/2},
	\label{eq:Eq.16}
\end{equation}

\noindent
which indicates the second-order nature of the exceptional line.
Next, we apply a small perturbation to $g$ with $g = g_{\mathrm{EL}} + \epsilon$ and $\kappa = \kappa_{\mathrm{EP}}$.
Following the same procedure we obtain
\begin{equation}
	\Delta E = \sqrt{2(F_{\mathrm{EL}} - G)}\epsilon^{1/2}\, \textrm{and} \,
	\Delta E = \sqrt{2g_{\mathrm{EL}}}\epsilon^{1/2},
	\label{eq:Eq.17}
\end{equation}

\noindent
confirming the second-order nature of the exceptional line.

\begin{figure}[!b]
	\begin{center}
		\includegraphics[width=0.5 \linewidth]{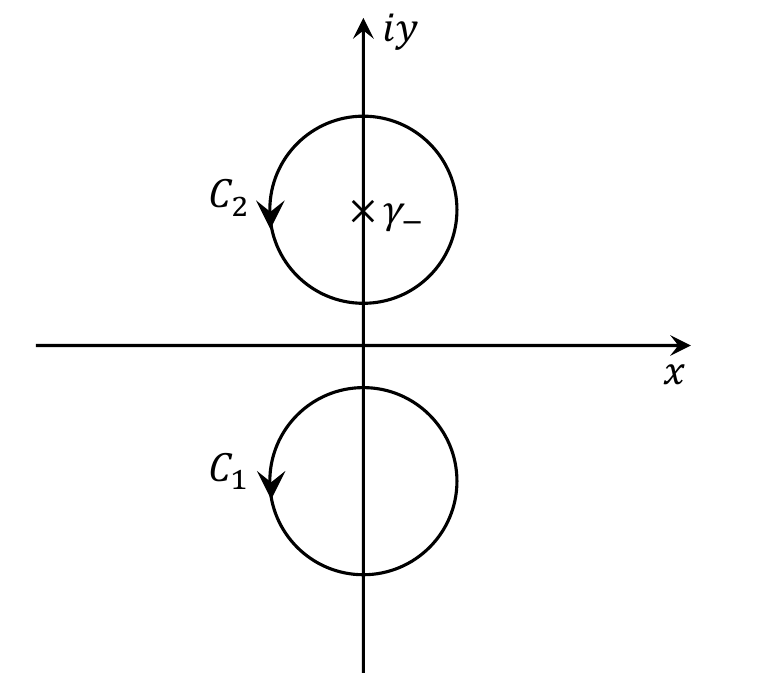} 
		\caption{
			Complex plane representation for calculation of the winding numbers.
			Circles $C_{1}$ and $C_{2}$ are closed contours oriented clockwise.
			The pole of  $E'_{\pm}$ is denoted by $\gamma_{-}$ in the complex plane $z = x+iy$.
		}
	\label{fig:winding_number}
	\end{center}
\end{figure}

\section{Calculation of the Winding Number}

The winding number is a topological invariant that quantifies the number of times that a function encircles a singularity along a closed contour~\cite{Ding2022}.  
To compute the winding number of the eigenvalues, we start from Eq.~(2) in the main text, rewritten as  
\begin{equation}
	E_{\pm} = \left( \frac{\omega_{\textrm{c}} + \omega_{\textrm{a}}}{2} - i \frac{\kappa + \gamma}{2} \right) \pm \sqrt{\left( \frac{\omega_{\textrm{c}} - \omega_{\textrm{a}}}{2} - i \frac{\kappa - \gamma}{2} \right)^{2} + g^{2} }.
	\label{eq:Eq.18}
\end{equation}

\noindent 
Substituting $\Delta_{\textrm{ca}} = \omega_\textrm{c} - \omega_\textrm{a}$ and defining $\gamma_+ = (\kappa + \gamma)/2$ and $\gamma_- = (\kappa - \gamma)/2$, we obtain  
\begin{equation}
	E_{\pm} = \left( \frac{\Delta_{\textrm{ca}} + 2\omega_{\textrm{a}}}{2} - i \gamma_+ \right) \pm \sqrt{\left( \frac{\Delta_{\textrm{ca}}}{2} - i \gamma_- \right)^{2} + g^{2} }.
	\label{eq:Eq.19}
\end{equation}

\noindent 
Further, letting $x = \Delta_{\textrm{ca}}/2$ and $y = g$, $E_{\pm}$ is given by
\begin{equation}
	E_{\pm} = \left( x + \omega_{\textrm{a}} - i \gamma_+ \right) \pm \sqrt{\left( x - i \gamma_- \right)^2 + y^2 }.
	\label{eq:Eq.20}
\end{equation}

\noindent 
To properly account for the multi-valued nature of the eigenvalues, we define modified eigenvalues $E'_{\pm}$ as  
\begin{align}
	E'_{\pm} & = E_{\pm} - \left( x +   \omega_{\textrm{a}}  - i \gamma_+ \right)\\ 
			 & = \pm \sqrt{\left( x - i \gamma_- \right)^{2} + y^{2} }.
	\label{eq:Eq.21}
\end{align}

\noindent 
Introducing a complex variable $z = x+iy$ and its conjugate $\bar{z} = x-iy$, we rewrite $E'_{\pm}$ as
\begin{equation}
	E'_{\pm} = \pm \sqrt{ (z - i \gamma_{-}) (\bar{z} - i \gamma_{-}) }.
	\label{eq:Eq.22}
\end{equation}

\noindent 
We then define a winding number $\mathcal{W}_{+}$ along a closed contour $C_{z}$ 
\begin{equation}
	\mathcal{W}_{+} = \frac{1}{2\pi i} \oint_{C_{z}} \frac{d}{dz} \log(E'_{+}) \, dz.
	\label{eq:Eq.23}
\end{equation}

\noindent 
Substituting $E'_{+}$ and differentiating, we obtain  
\begin{equation}
	\mathcal{W}_+ = \frac{1}{4\pi i} \oint_{C_z} \frac{1}{(z - i \gamma_-)} \, dz.
	\label{eq:Eq.24}
\end{equation}

\noindent Evaluating $\mathcal{W}_+$ along the contours $C_1$ and $C_2$, which respectively exclude and enclose the pole at $z = i\gamma_-$, yields  
\begin{equation}
	\mathcal{W}_+ = 
	\begin{cases} 
		0, & C_1 \\ 
		+\frac{1}{2}. & C_2 \\ 
	\end{cases}
	\label{eq:Eq.25}
\end{equation}

\noindent 
Similarly, computing $\mathcal{W}_-$ results in $\mathcal{W}_- = 1/2$.  
Consequently, the total winding number is given by  
\begin{equation}
	\mathcal{W} = \mathcal{W}_{+} + \mathcal{W}_{-} = +1.
	\label{eq:Eq.26}
\end{equation}

\section{References}
\bibliography{bibliography}